\documentclass[preprint,12pt,authoryear]{elsarticle}

\usepackage{amssymb}
\usepackage{amsmath}
\usepackage{xurl} 
\usepackage{hyperref}
\usepackage{float}

\usepackage{verbatim} 

\newcommand{\RootFile}{MAINARTICLE.tex}

\newcommand{\PrintWordcountReport}{%
  \immediate\write18{texcount -inc -merge -sub=section -utf8 "\RootFile" > "\jobname.wc"}%
  \clearpage
  \section*{Word count by section}
  \verbatiminput{\jobname.wc}%
}

\journal{Computers in Human Behavior}

\begin{document}

\begin{frontmatter}



\title{Metacognition and Confidence Dynamics in Advice Taking from Generative AI} 


\author{Clara Colombatto\corref{cor1}} 

\affiliation{organization={Department of Psychology, University of Waterloo},country={Canada}}

\author{Sean Rintel} 

\affiliation{organization={Microsoft Research},city={Cambridge},
           country={United Kingdom}}

\author{Lev Tankelevitch} 
\affiliation{organization={Microsoft Research},
           city={Cambridge},
           country={United Kingdom}}

\cortext[cor1]{Correspondence: Clara Colombatto, 200 University Ave W, Waterloo, ON N2L 3G1; clara.colombatto@uwaterloo.ca}

\begin{abstract}

Generative Artificial Intelligence (GenAI) can aid humans in a wide range of tasks, but its effectiveness critically depends on users being able to evaluate the accuracy of GenAI outputs and their own expertise. Here we asked how confidence in self and GenAI contributes to decisions to seek and rely on advice from GenAI (‘prospective confidence’), and how advice-taking in turn shapes this confidence (‘retrospective confidence’). In a novel paradigm involving text generation, participants formulated plans for events, and could request advice from a GenAI (Study 1; N=200) or were randomly assigned to receive advice (Study 2; N=300), which they could rely on or ignore. Advice requests in Study 1 were related to higher prospective confidence in GenAI and lower confidence in self. Advice-seekers showed increased retrospective confidence in GenAI, while those who declined advice showed increased confidence in self. Random assignment in Study 2 revealed that advice exposure increases confidence in GenAI and in self, suggesting that GenAI advice-taking causally boosts retrospective confidence. These results were mirrored in advice reliance, operationalised as the textual similarity between GenAI advice and participants’ responses, with reliance associated with increased retrospective confidence in both GenAI and self. Critically, participants who chose to obtain/rely on advice provided more detailed responses (likely due to the output’s verbosity), but failed to check the output thoroughly, missing key information. These findings underscore a key role for confidence in interactions with GenAI, shaped by both prior beliefs about oneself and the reliability of AI, and context-dependent exposure to advice.
\end{abstract}





\end{frontmatter}


\pagebreak
\section{Introduction}
\label{sec:intro}
One of the most prominent technological advances of our times is Generative Artificial Intelligence (GenAI), which uses generative models trained on vast amounts of data to produce new text, images, and other kinds of outputs. Large Language Models (LLMs), in particular, are able to generate coherent and contextually appropriate text, and can thus support humans in a variety of tasks—from code generation and personalised assistance to problem-solving and creative writing \citep{song_impact_2024, noy_experimental_2023, dellacqua_navigating_2023, doshi_generative_2024}. Alongside technical sophistication, however, the effectiveness of LLMs in aiding humans ultimately also depends on users’ willingness and ability to appropriately rely on their support \citep{schemmer_appropriate_2023,passi_appropriate_2024}. Studies of GenAI use in domains such as programming \citep{prather_beyond_2025,prather_its_2023}, law \citep{choi_ai_2023}, copywriting \citep{chen_large_2023}, data science \citep{gu_how_2023}, education \citep{zhai_effects_2024}, and design \citep{gmeiner_exploring_2023} suggest that users may over-rely on the technology, inappropriately accepting erroneous outputs. Realising the potential of LLMs to support humans thus requires a study of factors that shape reliance on these systems \citep{botvinick_realizing_2022}.

Early work on human use of automated systems demonstrated that users’ decisions to rely on these systems depend on a variety of factors, including perceived properties of the system, users’ experience, and usage context \citep{madhavan_similarities_2007}. Ultimately, however, reliance on external systems requires that users weigh their beliefs about the accuracy of the system (i.e., their \textit{confidence in the system}) against metacognitive assessments of their own expertise (i.e., their \textit{self-confidence}) \citep{madhavan_similarities_2007,de_vries_effects_2003,lee_trust_1994}. 
For instance, users typically rely more on AI systems when they perceive the AI as highly accurate \citep{yin_understanding_2019,kahr_understanding_2024}, particularly in domains where their own expertise is limited \citep{castelo_task-dependent_2019}. Conversely, in tasks where users feel confident in their own knowledge or skills, they may be less inclined to seek or rely on AI advice, even if the system is demonstrably reliable \citep[for a review, see][]{jussupow_why_2020,jessupCloserLookHow2024}. Recent work in AI-assisted decision-making suggests that self-confidence may contribute to reliance more than confidence in the AI \citep{chong_human_2022}.

The role of confidence in reliance on automated systems is also supported by research on cognitive offloading—our tendency to rely on external tools to reduce cognitive demand \citep{risko_cognitive_2016}. When deciding whether to use external aids such as automated reminders or study notes, users weigh both their confidence in the external aid \citep{weis_using_2019}, and their confidence in themselves \citep{dunn_toward_2016, boldt_confidence_2019, hu_role_2019, scott_metacognition_2024}. Beyond tool use, confidence also guides reliance on advice from other people: decision makers are less likely to request advice when they are confident in their own abilities \citep{see_detrimental_2011,pescetelli_confidence_2021}. Conversely, decision makers are more likely to request advice when they are confident in the quality of the advice, either due to their beliefs about the expertise of the advisor \citep{carlebach_flexible_2023}, or due to the advisor’s communicated or perceived confidence \citep{sniezek_trust_2001,van_swol_factors_2005,pulford_social_2018,coucke_action-based_2024,colombatto_illusions_2023}.

While past work has demonstrated a role for confidence in both the self and in external systems in reliance on AI and human advisors, it remains unclear how these factors may contribute to reliance on GenAI \citep{tankelevitch_metacognitive_2024}. While GenAI may be conceived as another form of external support, it is also unique for key reasons. Relative to commonly available AI technologies such as recommender systems or navigation apps, GenAI systems exhibit greater flexibility in their input/output space, especially considering their probabilistic (rather than deterministic) output, and greater originality in that they can generate novel content \citep{schellaert_your_2023}. GenAI systems also allow for greater generality in that they are suitable to a diverse range of tasks, and are both useful for and used by the general population across a novel range of generative tasks \citep{handa_which_2025,schellaert_your_2023}—from programming, to writing or planning. In these types of tasks, advice reliance is also more nuanced, as users can choose to integrate some but not all the advice, allowing for more continuous metrics of reliance beyond those in the discrete and binary decision-making tasks commonly studied in prior research \citep{chong_human_2022, kahr_understanding_2024,jessupCloserLookHow2024}. Further, these novel task contexts entail complexity, and often subjectivity, in evaluations of performance, giving rise to a wide range of user confidence levels in both themselves and in the performance of GenAI systems \citep{simkute_ironies_2024,tankelevitch_metacognitive_2024}. These unique characteristics of GenAI systems thus raise important questions regarding the role of confidence in reliance on GenAI, particularly in generative tasks for which such systems are being commonly used. 

The uniqueness of GenAI in terms of its extensive and flexible outputs, also raises questions about how users evaluate its output, and how this influences their confidence in the systems. Indeed, confidence can be operationalized as \textit{prospective}—the confidence that users feel before a task, which shapes decisions to seek AI advice; but also \textit{retrospective}—the confidence users report after a task, shaped by their experiences and outcomes. In this sense, the relationship between confidence and reliance may be bi-directional, such that prospective confidence in the system determines reliance on its output, but output evaluation may in turn influence retrospective confidence in the system. In fact, reliance on external advice may even impact users’ own confidence: studies of internet use have shown that relying on the internet for information retrieval inflates users’ confidence in their own knowledge \citep{fisher_searching_2015,dunn_distributed_2021,eliseev_understanding_2023}. We thus considered the possibility that confidence may shape reliance in different ways—with prospective confidence shaping decisions to take AI advice, and advice in turn influencing retrospective estimates in responses \citep{fernandes_ai_2024}. 

We report two preregistered experiments investigating the role of prospective and retrospective confidence in advice-taking from a GenAI system, in the context of a generative task typical of real-world GenAI use. We designed a novel experimental paradigm in which participants were asked to formulate plans for different events and were given the opportunity to incorporate advice from a GenAI system. In Study 1, participants were given the choice to request or decline advice from the GenAI, while in Study 2, participants were randomly assigned to receive advice or not. By examining how prospective confidence shapes advice-taking from GenAI, and how advice-taking in turn affects retrospective confidence, this research contributes to a deeper understanding of metacognitive mechanisms that support effective and calibrated trust in AI.

\section{Study 1: Confidence and GenAI Support}\label{sec:study1}

To investigate the role of confidence on advice-taking from a GenAI system, we designed an event planning task where participants formulated plans for different events and were offered advice from a GenAI system. We hypothesized that requests for advice and reliance on the GenAI output would depend on participants’ prospective confidence in their own ability (with high confidence leading to low requests and reliance), as well as in the capability and usability of the GenAI system (with high confidence leading to high requests and reliance). We also hypothesized that reliance on the GenAI output would in turn increase retrospective confidence in the GenAI system.

\subsection{Method}\label{subsec:method1}

The methods and analyses for this study were preregistered and can be accessed at \url{https://aspredicted.org/568n-9qrr.pdf}. Anonymized raw data and analysis code are openly available on the Open Science Framework (OSF) website at \url{https://osf.io/nxpb6/?view_only=29a813136da24784b8f45bca0a9d8bae}. All experimental methods and procedures were approved by a University of Waterloo Research Ethics Board (Protocol \#46224), and all participants provided informed consent. 

\subsubsection{Participants}\label{subsubsec:participants1}

A total of 200 participants were recruited through Prolific (\url{www.prolific.com}) \citep{palan_prolificacsubject_2018} in exchange for monetary compensation, in February and March 2024. This preregistered sample size was chosen before data collection began based on an a priori power analysis, which revealed that a sample of 193 participants would be sufficient to achieve 80\% power to detect a weak to moderate correlation (r=0.20) with an alpha level of 0.05. Recruitment was stratified by age and gender categories to obtain a sample representative of the US population. In particular, data were collected for males and females across seven age groups (18-24, 25-34, 35-44, 45-54, 55-64, 65-74, over 75), with the number of participants in each of these 14 categories determined according to the corresponding representation in the population based on US census data.

To be eligible to take part in the study, participants were required to be fluent in English, have an approval rate of 95-100\%, be located in the U.S., and use a laptop or desktop (as opposed to a tablet or phone). Participants were excluded from our dataset per our pre-registered criteria if they (1) participated more than once, as determined via their platform ID (N=0); (2) failed to select the correct response in an attention check at the end of the experiment (N=11); (3) reported having encountered problems (as assessed via a question in debriefing; N=0); (4) failed to answer any of our open-ended questions sensibly (N=0); and (5) failed to enter text for two or more of the four scenarios (N=13). These participants were excluded and replaced until our target sample size was reached (N=200, 100 women, 94 men, 5 non-binary, 1 preferred not to answer; mean [M] age=46.55, SD=18.47; see Supplementary Information for more details on participant demographics). At the trial level, we excluded scenarios where participants (1) submitted their final response in less than 60 seconds (N=0); and (2) did not enter any text in the response box (N=5), for a total of N=795 remaining trials.

\subsubsection{Design}\label{subsubsec:design1}

Each participant was asked to formulate plans for four events; two concerned work-related occasions, namely a team-building retreat and an office recruitment event; and the other two concerned personal activities, namely a weekend camping trip and a dinner party. Participants were randomly assigned to receive advice in bullet points or in full prose (for exploratory purposes of output format; not reported here). As we were interested specifically in the role of confidence on advice-taking and reliance on GenAI, we sought to control for differences in people’s prompting abilities. As such, we used a single, consistent prompt that was submitted to ChatGPT (model gpt-4-2024-02-15-preview) during the experimental design stage, and precluded participants from being able to edit the prompt. All participants were shown the same AI-generated output. A key planning step was manually removed from each output for analyses of participants’ accuracy. See Supplementary Information for the full text of each prompt shown to participants and the corresponding GenAI output.

\subsubsection{Procedure}\label{subsubsec:procedure1}

Participants were redirected to a website with custom software written in HTML, CSS, and JavaScript, using the jsPsych library \citep{de_leeuw_jspsych_2015} and the JATOS server \citep{lange_just_2015}. All participants provided their informed consent, and their browser window was put in full-screen mode at the beginning of the experiment. Participants were then told that they would be asked about different types of events, one at a time, and for each, they would be asked to “come up with a list of steps you would take to organise each event”. They were also told that “if you’d like help with the plan, you will also have the opportunity to obtain information from a Generative AI system (GenAI), similar to ChatGPT. This is a sophisticated and powerful computer program that has been trained on vast amounts of text data, enabling it to answer questions and generate coherent text.” Importantly, we noted that “There is no requirement to look at and/or use the GenAI system’s information; you can rely on it as much or as little as you want. GenAI systems can make mistakes. Consider checking important information.” To encourage accuracy and compliance, they were also informed that the most complete plan for each scenario would be awarded a USD \$10 bonus; since all participants completed four scenarios, each participant had the opportunity to win up to USD \$40.

Participants then completed each of the four scenarios, in a random order. Each scenario was first introduced with a description (e.g., “Imagine you need to organise a one-day retreat with your co-workers at a school that you work in to foster team-building.”), followed by three questions probing participants’ prospective confidence, including confidence in themselves, confidence in the GenAI, and confidence in GenAI usability (see \autoref{table1_confquestions}). These questions were displayed in a random order on separate screens, and all were to be answered on a scale from 1 to 7 , with endpoints labelled “Not confident at all” and “Very confident”. 

\renewcommand{\arraystretch}{1.5}
\begin{table}[t]
\centering
\footnotesize
\begin{tabular}{p{0.1\textwidth} p{0.4\textwidth} p{0.4\textwidth} }
   & \textbf{Prospective Confidence} & \textbf{Retrospective Confidence} \\ \hline
  \textbf{Self} & \textit{How confident are you in planning this event?} & \textit{How confident were you in planning this event?} \\ \hline
  \textbf{GenAI} & \textit{How confident are you in GenAI systems’ information for planning this event? If you don’t have any prior experience with GenAI systems, provide your best guess.} & \textit{How confident were you in the GenAI system’s information for planning this event? If you didn’t use the GenAI system, give us your best guess imagining you had used it.} \\ \hline
  \textbf{Usability} & \textit{How confident are you in working with GenAI systems on planning this event? If you don’t have any prior experience with GenAI systems, provide your best guess.} & \textit{How confident were you in working with the GenAI system on planning this event? If you didn’t use the GenAI system, give us your best guess imagining you had used it.} \\ \hline
\end{tabular}
\caption{Measures of Confidence in a Generative Task. Participants were asked to indicate their confidence in their own ability, the capability of the GenAI system, and the usability of the GenAI system, both prior to each trial (prospective confidence) and just after submitting each response (retrospective confidence). These questions were displayed in a randomized order, and were all answered on a scale from 1 (“Not confident at all”) to 7 (“Very confident”).}\label{table1_confquestions}
\end{table}
\renewcommand{\arraystretch}{1}

\begin{figure}[t]
\centering
\includegraphics [width=\linewidth]{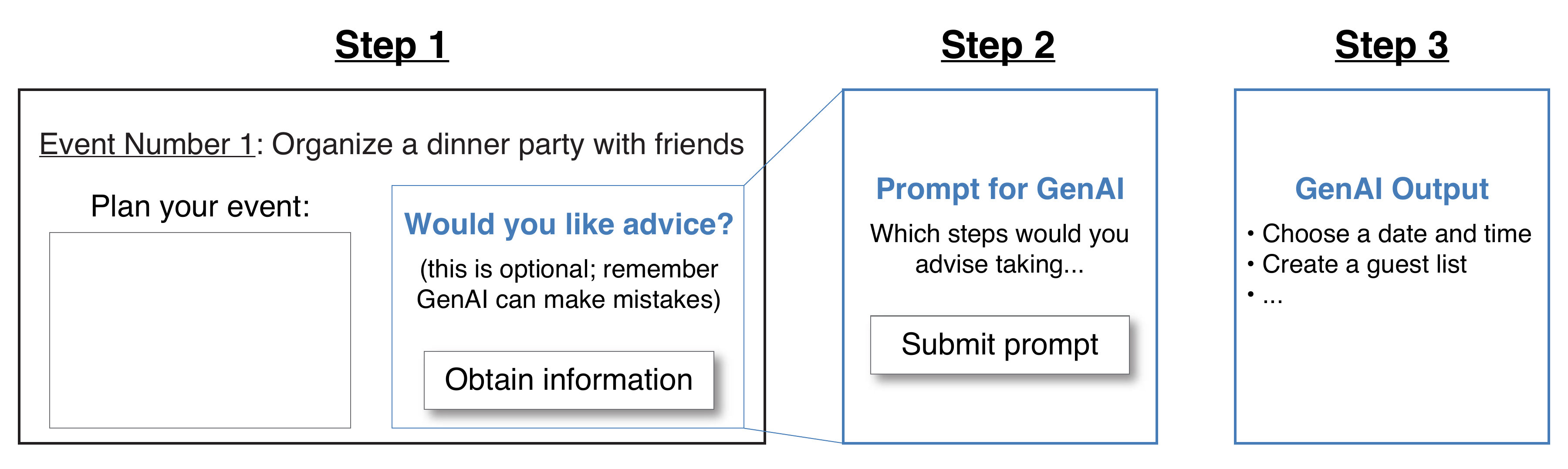}
\caption{A Novel Task to Measure Reliance on GenAI. Participants were asked to formulate a plan for an upcoming event, and had the opportunity to request advice from a GenAI system. Each participant completed a total of four events (work retreat, office recruitment, camping trip, dinner party) in a randomized order. Note that text has been simplified for visualization purposes.}\label{fig1_task}
\end{figure}

After indicating their prospective confidence, participants were shown the main task screen (\autoref{fig1_task}). On the left side of this screen was a textbox, where participants could type their plan (with the title \textit{“Which steps would you take to plan this event?”}, and a pre-filled placeholder text \textit{“Start typing here”}); and on the right side was the GenAI information, namely the option to receive advice (\textit{“Would you like to obtain information from a GenAI system? (There is no requirement to look at and/or use the GenAI system information; you can rely on it as much or as little as you want. GenAI systems can make mistakes. Consider checking important information.)”}) along with a button to \textit{“Obtain information from GenAI system”}. If participants selected to obtain information, they then viewed a pre-formulated prompt (e.g., \textit{“Which steps would you advise taking to organise a one-day retreat with your co-workers at a school that you work in to foster team-building?”}), along with a button to \textit{“Submit prompt”}, which then led to the output (\textit{“GenAI system output”}). At first, this was just a placeholder (\textit{“Generating…”}), followed by the information after 2s (e.g., \textit{“When organizing a one-day retreat for team-building at your school, it’s essential to begin by clearly defining the objectives you aim to achieve….”}). Throughout the task, there was a button at the bottom of the screen to ‘Continue’; this could be selected at any point during the task, so participants could proceed to the next screen, even without requesting the advice.

After submitting their plans, participants answered three questions probing their retrospective confidence. These mirrored the prospective confidence questions, and included confidence in themselves, confidence in the GenAI, and confidence in GenAI usability (\autoref{table1_confquestions}). For exploratory purposes, they were asked two additional questions about the provided prompt (\textit{“How good do you think the provided prompt was for planning this event?”}, to be answered on a 1-7 scale from \textit{“Not at all good”} to \textit{“Very good”}; and \textit{“What question would you have asked the Gen AI system to obtain information for planning this event? Write the prompt you would have submitted below”}). 

At the end of the study, participants were asked several questions (\autoref{table2_generalquestions}), concerning general assessment of their own abilities, their trust in the GenAI systems and their usability, and their experience with GenAI. Finally, they were asked questions about their demographics, including gender, age, education, area of study, employment status, and area of employment; open-ended questions asking if they experienced any technical issues or interruptions; and a multiple-choice question probing their attention (\textit{“Which of the following scenarios was NOT one of the events you were asked to plan today:”} [A graduation party - A weekend camping trip - A dinner party - An office recruitment event - A team-building retreat]). 

\renewcommand{\arraystretch}{1.5}
\begin{table}[H]
\footnotesize
\centering
\begin{tabular}{p{0.2\textwidth} p{0.4\textwidth} p{0.3\textwidth}}
  \textbf{Theme} & \textbf{Question} & \textbf{Response options} \\ \hline 
  Confidence in Self & \textit{How good do you think you are at planning events?} & 1 (“Not at all good”) to 7 (“Very good”) \\
   & \textit{How good do you think you are at working with Generative AI systems?} & 1 (“Not at all good”) to 7 (“Very good”) \\ \hline
 Confidence in GenAI Systems & \textit{How good do you think Generative AI systems are at planning events?} & 1 (“Not at all good”) to 7 (“Very good”) \\
   & \textit{To what extent do you think Generative AI systems can provide accurate and helpful suggestions in general?} & 1 (“Not at all helpful”) to 7 (“Very helpful”) \\  \hline
Confidence in GenAI Usability & \textit{To what extent do you think the Generative AI system helped you formulate better plans in the scenarios you worked on in this experiment?} & 1 (“Not at all helpful”) to 7 (“Very helpful”) \\
   & \textit{How comfortable do you feel relying on Generative AI systems for important tasks or decisions?} & 1 (“Not at all comfortable”) to 7 (“Very comfortable”) \\  
   & \textit{Do you feel in control when using Generative AI systems to assist with event planning decisions?} & 1 (“Not at all in control”) to 7 (“Very in control”) \\  
   & \textit{How important is it for you to have the ability to modify the suggestions of Generative AI systems?} & 1 (“Not at all important”) to 7 (“Very important”) \\  \hline
Experience with GenAI & \textit{Had you heard of Generative AI (GenAI) systems prior to this experiment?} & Yes / No \\
   & [If yes] \textit{Have you used Generative AI systems in the past?} & Yes / No \\  
   & [If yes] \textit{How often do you use GenAI systems (e.g., ChatGPT, Bing Copilot)?} & More than once a day / About once a day / About once a week / About once every two weeks / Less than once a month / About once a month \\  \hline
\end{tabular}
\caption{Measures of General Beliefs and Attitudes. At the end of the experiment, participants were asked several questions about their beliefs about their own abilities, their perceptions of GenAI systems and their usability, and their experience with GenAI systems.}\label{table2_generalquestions}
\end{table}
\renewcommand{\arraystretch}{1}

\subsubsection{Overview of Analyses}\label{subsubsec:overview1}

An initial analysis of the relationship between the three measures of confidence revealed that confidence in the GenAI and confidence in GenAI usability were highly correlated (prospective: r=.77, p<.001; retrospective: r=.82, p<.001). Due to the high collinearity between these two measures, we collapsed them in the analyses—and we also confirmed that including either of these measures instead does not alter the results in any meaningful way. Unless otherwise noted, all analyses described below include participant and event as random intercepts, and confidence variables (i.e. prospective and retrospective confidence in self, prospective and retrospective confidence in GenAI) were z-scored.

\paragraph{Advice-Taking} To examine the role of confidence in advice-taking from GenAI, we analysed the relationship between prospective confidence and advice-taking, with the latter measured in two distinct ways: (1) advice requests, namely whether participants choose to request advice from the GenAI system (in response to the question \textit{“Would you like to obtain information from a GenAI system?”}); and (2) advice reliance, namely the extent to which they incorporate the output into their responses, operationalised as the cosine similarity between the GenAI output and the participants’ plan. This resulted in the following analyses: (1) a binomial regression predicting advice requests (Yes/No) from prospective confidence in self, prospective confidence in the GenAI, and their interaction (reported in Section \ref{subsubsec:results1_1}, \textit{Advice Requests}); and (2) a linear regression predicting advice reliance (from 0 to 1, on trials in which advice was requested) from prospective confidence in self, prospective confidence in the GenAI, and their interaction (reported in Section \ref{subsubsec:results1_2}, \textit{Advice Reliance}).

Next, we examined the impact of advice-taking and prospective confidence on retrospective measures of confidence as measured at the end of each trial, for both confidence in self and confidence in the GenAI (see \autoref{table1_confquestions}). This resulted in the following analyses (reported in Section \ref{subsubsec:results1_3}, \textit{From Prospective to Retrospective Confidence}): (3a) a linear regression predicting retrospective confidence in self from prospective confidence in self, advice request (Yes/No), and their interaction; (3b) a linear regression predicting retrospective confidence in GenAI from prospective confidence in GenAI, advice request (Yes/No), and their interaction; (4a) a linear regression predicting retrospective confidence in self from prospective confidence in self, advice reliance (from 0 to 1), and their interaction; and (4b) a linear regression predicting retrospective confidence in GenAI from prospective confidence in GenAI, advice reliance (from 0 to 1), and their interaction. 

\paragraph{Accuracy} Next, we examined participants’ performance in the event planning task, with accuracy operationalised in two ways: (1) response verification, namely whether participants included in their plan a critical step that was removed from the advice of the GenAI; and (2) response completeness, namely how many steps from an ‘ideal’ response were mentioned. These scores were obtained via ChatGPT (model gpt-4, api version 2024-02-15-preview) using chain-of-thought prompting. For example, for response completeness, the model was first asked to assign a similarity score to each step by comparing individual steps in participants’ plans with the corresponding steps in the model response. Then, a cumulative completeness score was computed based on how many steps from the model response were represented in participants' plans. See Supplementary Information for full details on scoring for both measures. This resulted in the following analyses (reported in Section \ref{subsubsec:results1_4}, \textit{Metacognitive Calibration}): for response verification, (5a) a binomial regression predicting the likelihood of including the omitted piece of information from prospective confidence in self, advice request (Yes/No), and their interaction; (5b) a binomial regression predicting the likelihood of including the omitted piece of information from prospective confidence in GenAI, advice request (Yes/No), and their interaction; and for response completeness, (6a) a regression predicting the number of steps mentioned in the response from prospective confidence in self, advice request (Yes/No), and their interaction; and (6b) a regression predicting the number of steps mentioned in the response from prospective confidence in GenAI, advice request (Yes/No), and their interaction.

\paragraph{Individual Differences} In a series of secondary analyses, we explored possible determinants of advice request and reliance based on participants’ self-reported attitudes and beliefs (see Supplementary Information for response distributions), in analyses of (7) attitudes towards GenAI (e.g., perceived planning capability, helpfulness in the planning task or in general, comfort relying on and working with GenAI), (8) usage frequency; and (9) demographic characteristics (gender, age, education, employment).

\subsection{Results}\label{subsec:results1}

\subsubsection{Advice Requests}\label{subsubsec:results1_1}
\begin{figure}[H]
\centering
 \includegraphics [width=\linewidth]{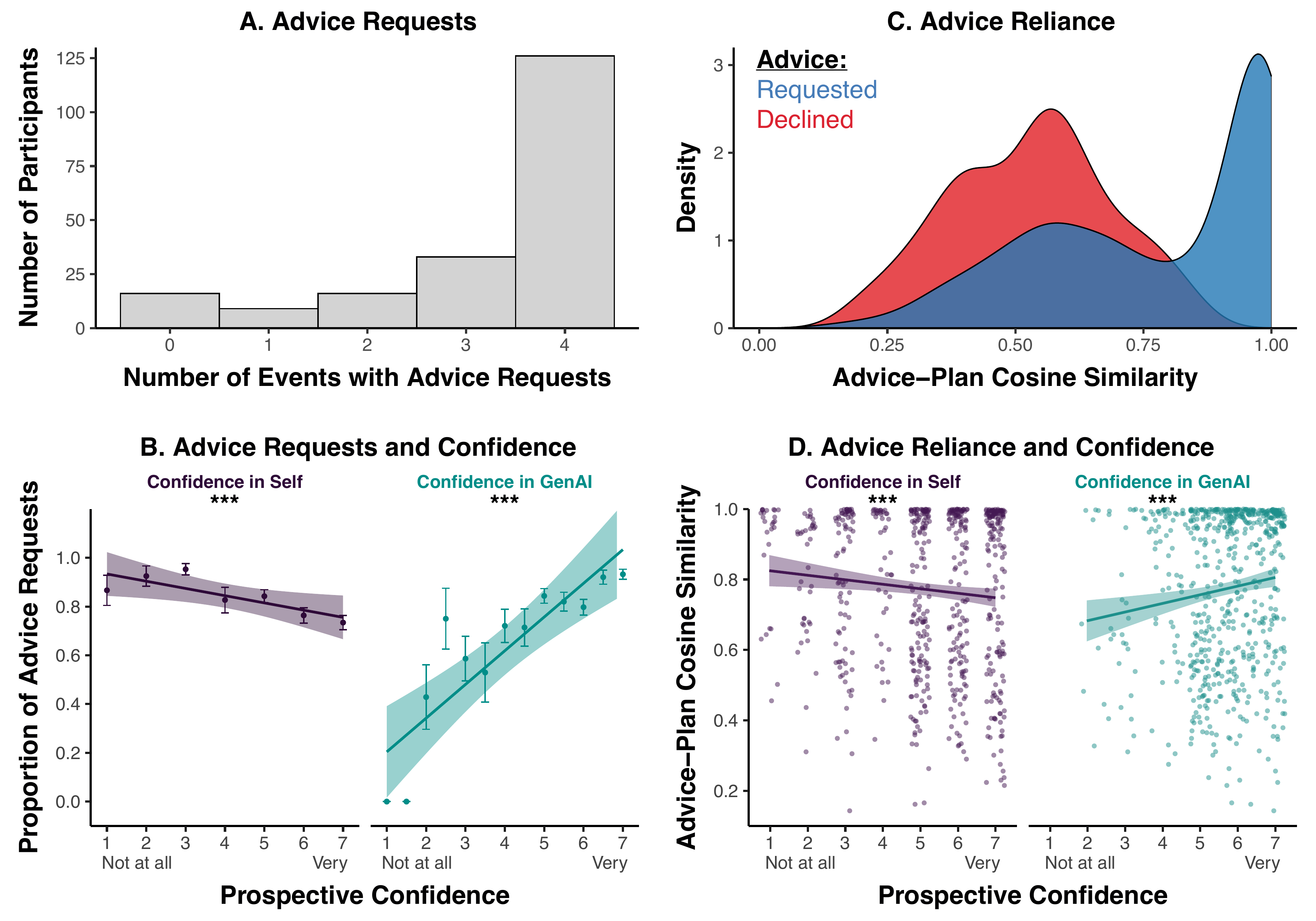}
\caption{Relationship Between Confidence and Advice-Taking. (A) Number of subjects who requested advice on no trials (0), some (1-3), or all trials (4). (B) Participants who were more confident in themselves were less likely to request advice; those who were more confident in GenAI were instead more likely to request advice. Points correspond to averages across all participants for each confidence value (from 1 to 7), and error bars correspond to standard errors; for this and subsequent plots, lines correspond to best-fit linear regression line, and shaded bands represent 95\% confidence intervals. (C) Density of GenAI advice-participants' plan similarity, separately for participants who requested (blue) or declined (red) advice. (D) Participants who were more confident in themselves were less likely to rely on the advice (measured as the cosine similarity between the advice provided by the GenAI and the plan submitted by participants); those who were more confident in the GenAI were instead more likely to rely on the advice. Jittered points correspond to single trials where participants requested advice (with confidence in GenAI averaged across two questions; note that, for this and subsequent plots, points corresponding to individual trials were jittered to improve the visibility of overlapping observations within each confidence level, while still preserving clear separation between confidence levels).}\label{fig_study1results1}
\end{figure}

Participants decided to request advice on most trials, with advice requests submitted on 81\% of events; participants requested advice on 3.22 out of 4 events on average; 126 out of the 200 participants requested advice on all four events, while only 16 participants never did (\autoref{fig_study1results1}A). To investigate the role of confidence on decisions to request advice, we ran a preregistered binomial regression predicting advice requests (Yes/No) from the two prospective confidence ratings (self, GenAI) and their interaction. Participants who were more confident in the GenAI were more likely to request advice (main effect of confidence in GenAI: B=1.64, SE=0.27, z=6.15, p<.001; \autoref{fig_study1results1}B, right panel). Those who were more confident in their own planning abilities instead were less likely to request advice (main effect of confidence in self: B=-1.68, SE=0.32, z=-5.17, p<.001; \autoref{fig_study1results1}B, left panel), with no interaction between these effects (p=.524). Decisions to request advice are thus associated with both lower confidence in one’s own abilities, and higher confidence in the GenAI’s abilities.

\subsubsection{Advice Reliance}\label{subsubsec:results1_2}
Next, we asked whether confidence may be associated with reliance on GenAI advice, above and beyond decisions to request advice. The extent of advice reliance was quantified as the cosine similarity (from 0 to 1) between the GenAI advice provided and the final submitted answer (\autoref{fig_study1results1}C). On trials where advice was requested, participants submitted plans that were highly similar to the GenAI output, with 53\% of these responses showing greater than 80\% similarity with the advice (compared to only 5\% for responses where advice was declined). Similarity was analysed via a linear model with the two prospective confidence ratings and their interaction as fixed effects; note that these analyses only include trials in which advice was requested (and by implication, only those participants who requested advice on these trials). Overall, advice reliance mirrored the results of advice requests: participants who were more confident in their own planning abilities relied on the advice to a lesser extent (main effect of confidence in self: B=-0.03, SE=0.01, t(377)=-4.71, p<.001; \autoref{fig_study1results1}D, left panel), while those who were more confident in the GenAI relied on the advice to a greater extent (main effect of confidence in GenAI: B=0.02, SE=0.01, t(544)=2.68, p=.008; \autoref{fig_study1results1}D, right panel), with no interaction between these measures of confidence (p=.112). Confidence thus plays a role not just in deciding whether to take GenAI advice, but also in how the advice is incorporated into a final decision: reliance on GenAI advice is associated with both lower confidence in one’s own abilities, and greater confidence in the GenAI’s abilities.

\subsubsection{From Prospective to Retrospective Confidence}\label{subsubsec:results1_3}

To examine how confidence estimates changed from before to after completing the plan on each trial, we ran two linear regressions predicting retrospective measures of confidence (in self and in GenAI) from the corresponding prospective measure, advice requests, and their interaction. Overall, participants who were more confident prior to the task were also more confident after the task, for both types of confidence (main effect of prospective confidence in self: B=0.55, SE=0.06, t(662)=8.73, p<.001; main effect of prospective confidence in GenAI: B=0.63, SE=0.05, t(777)=12.97, p<.001). Retrospective confidence was also overall related to decisions to take advice, with participants who declined advice being more confident in their own abilities (main effect of advice taking on retrospective confidence in self: B=-0.19, SE=0.06, t(622)=-3.04, p=.002), and participants who requested advice being overall more confident in the GenAI (main effect of advice taking: B=0.44, SE=0.08, t(738)=5.86, p<.001).

\begin{figure}[H]
\centering
 \includegraphics [width=\linewidth]{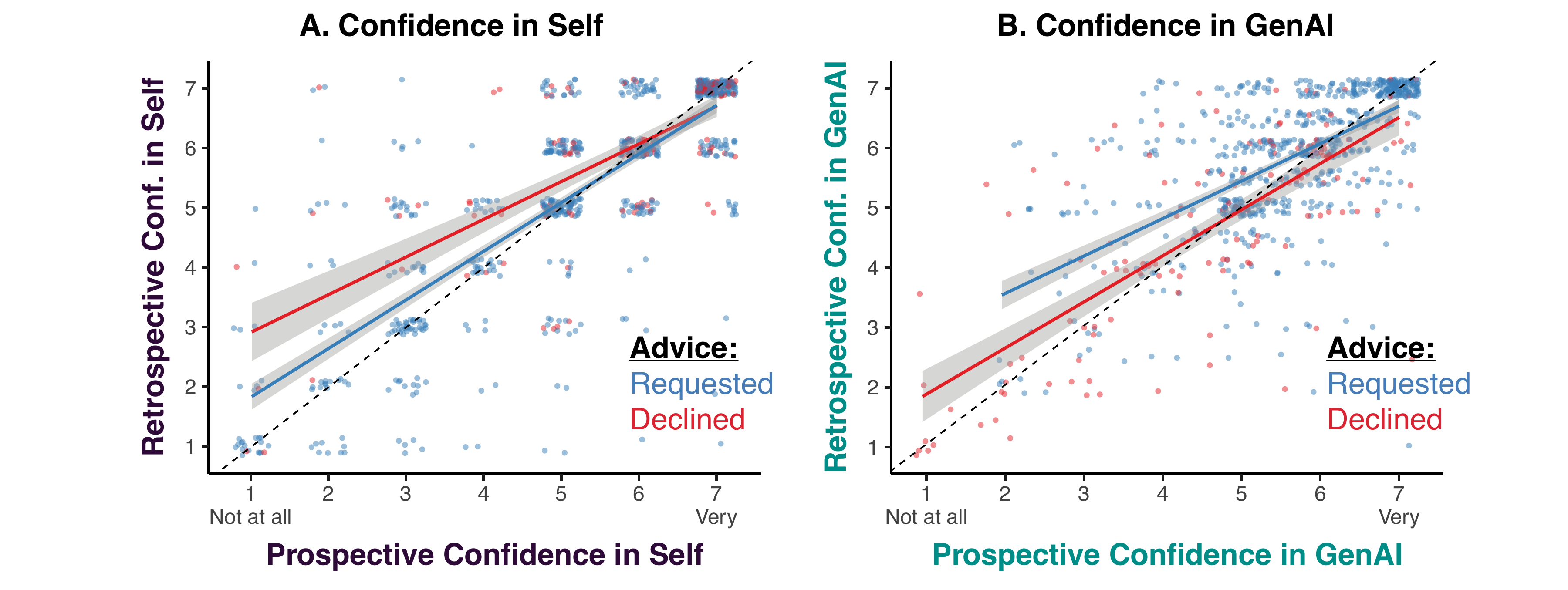}
\caption{Changes from Prospective to Retrospective Confidence in Study 1. (A) Confidence in self was overall highly consistent from before to after the task, but participants who declined advice showed a boost in confidence after the task. (B) Confidence in GenAI was overall highly consistent from before to after the task, but participants who requested advice showed a boost in confidence after the task. Jittered points correspond to single trials (with confidence in GenAI averaged across two questions).}\label{fig_study1results2}
\end{figure}

 Importantly, there was also an interaction between advice requests and prospective confidence. For confidence in self, participants who requested advice had a strong correlation between prospective and retrospective ratings (slope=0.75, SE=0.03, CI=[0.69, 0.80]), while this correlation was weaker in those who declined advice (slope=0.55, SE=0.06, CI=[0.43, 0.68]; interaction B=0.19, SE=0.07, t(723)=2.87, p=.004; \autoref{fig_study1results2}A). In other words, those who declined advice had a larger increase in self-confidence after the task, compared to those who requested advice. For confidence in GenAI, the correlation between prospective and retrospective ratings was stronger in participants who declined advice (slope=0.63, SE=0.05, CI=[0.53, 0.72]) compared to those who requested advice (slope=0.45, SE=0.03, CI=[0.39, 0.52]; interaction B=-0.17, SE=0.06, t(791)=-3.06, p=.002; \autoref{fig_study1results2}B). In other words, those who requested advice had a larger increase in confidence in GenAI, compared to those who declined advice. There was thus a boost in self-confidence after task completion, but more so in participants who declined advice; conversely, confidence in GenAI increased for participants who requested advice, but less so for those who declined advice.

These interactions between confidence and advice requests also held when considering advice reliance on trials where participants did request the advice. For confidence in self, there was a strong main effect of prospective confidence (B=0.48, SE=0.10, t(636)=4.90, p<.001) and no main effect of advice reliance (p=.743), but there was an interaction between these factors (B=0.31, SE=0.12, t(638)=2.68, p=.008): participants who relied more on the advice (reliance=1.0) had a strong correlation between prospective and retrospective ratings (slope=0.79, SE=0.04, CI=[0.72, 0.86]), while this relationship was weaker in those who relied partially on the advice (reliance=0.5: slope=0.64, SE=0.05, CI=[0.55, 0.73]) and those who did not rely on the advice at all (reliance=0.0; slope=0.48, SE=0.10, CI=[0.29, 0.67]). For confidence in GenAI, there was a main effect of prospective confidence (B=0.54, SE=0.11, t(630)=5.14, p<.001) and a main effect of advice reliance (B=0.59, SE=0.17, t(382)=3.49, p=.001); the interaction between these factors was weak (B=-0.13, SE=0.13, t(631)=-0.98, p=.329), although simple comparisons confirmed a numerical trend in the same direction as with advice requests: participants who relied more on the advice (reliance=1.0) had a positive correlation between prospective and retrospective ratings (slope=0.41, SE=0.05, CI=[0.33, 0.50]), and this relationship was stronger in those who relied only partially on the advice (reliance=0.5: slope=0.48, SE=0.05, CI=[0.39, 0.57]) and in those who did not rely on the advice at all (reliance=0.0; slope=0.54, SE=0.11, CI=[0.34, 0.75]). Additional analyses of the two measures of confidence in GenAI revealed that this weak interaction was due to a null interaction for confidence in the information provided by the GenAI (p=.917), but a stronger interaction for confidence in GenAI usability (B=-0.30, SE=0.14, t(630)=-2.16, p=.031). This modulation of confidence by advice reliance is thus consistent with the effects of advice requests, such that participants who declined, or requested but did not rely on the advice, showed increased self-confidence after the task, while participants who requested and relied on the advice showed increased confidence in the GenAI.

\subsubsection{Metacognitive Calibration}\label{subsubsec:results1_4}
\begin{figure}[H]
\centering
 \includegraphics [width=\linewidth]{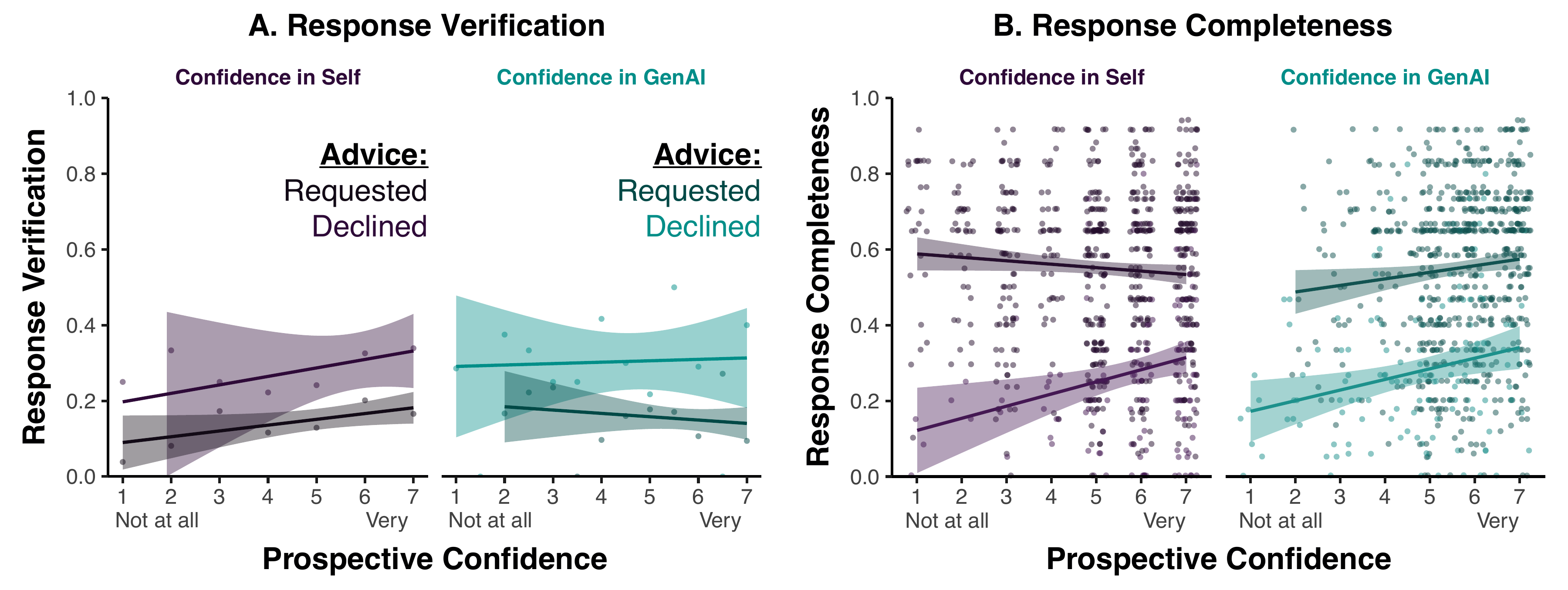}
\caption{Confidence Calibration in Study 1. (A) Participants who declined advice were more likely to include in their plan a piece of information that was removed from the advice. Confidence in both self and GenAI was overall unrelated to this measure of response verification (perhaps due to floor effects as per the overall low response verification rates). Points correspond to averages across all participants for each confidence value (from 1 to 7). (B) Participants who requested advice provided more complete responses (perhaps due to high reliance on the advice, which was highly detailed). Participants who were more confident in themselves provided more complete responses, but only when they declined advice; those who requested the advice instead showed a negative relationship between accuracy and confidence, suggesting that advice from GenAI disrupts metacognitive calibration. Jittered points correspond to single trials (with confidence in GenAI averaged across two questions).}\label{fig_study1results3}
\end{figure}

In a set of secondary analyses, we investigated participants’ accuracy in the task and its relationship with their confidence. First, we considered response verification, namely whether participants included in their plan a critical step that was removed from the advice of the GenAI, unbeknownst to them (see Supplementary Information for full details on the omitted steps). Each plan was scored by submitting the participant’s final response to ChatGPT (model gpt-4-2024-02-15-preview) and asking whether it explicitly mentioned the steps (see Supplementary Information for full details on the scoring procedure). Participants who did not request advice were more likely to verify the output, i.e. to include in their plan the critical step compared to participants who did receive advice with the information taken out (30.46 vs. 15.22\% of plans that included the information; X2(1)=18.16, p<.001). We next investigated the relationship between prospective confidence and response verification, in a model predicting the likelihood of including information omitted from the output from prospective confidence in self and advice request. Confidence was unrelated to response verification (p=.836), and while participants who declined advice were more likely to include the omitted information (B=-0.78, SE=0.29, z=-2.71, p=.007), this was not associated with confidence in self (p=.771; \autoref{fig_study1results3}A). Similarly, response verification was unrelated to confidence in the GenAI (p=. 845), and while participants who declined advice were again more likely to include the omitted information (B=-0.75, SE=0.29, z=-2.56, p=.011), confidence did not mediate this effect (p=.667). This suggests that response verification was strongly dependent on advice-taking, with participants who requested advice being less likely to incorporate omitted information, perhaps because they were not verifying the output closely. On the other hand, verification was largely unrelated to participants’ confidence, suggesting poor calibration—although we interpret this result with caution given the overall low rates of inclusion for this omitted piece of information (see \autoref{fig_study1results3}A). 

As an additional metric of accuracy, we computed the overall response completeness, by retrieving via ChatGPT the number of steps included in each plan out of all steps included in a model answer (see Supplementary Information for full details on response scoring). Participants who did not request advice provided overall less detailed plans (mean \% points included=27.39\%, SE=2.38) compared to those who did request advice (mean score=55.02, SE=1.63). Confidence in self correlated with response completeness for participants who declined the advice: there was a main effect of advice request (B=23.95, SE=2.01, t(785)=11.90, p<.001; \autoref{fig_study1results3}B) and no main effect of confidence in self (p=.292), but importantly there was an interaction between these factors (B=-4.06, SE=2.03, t(759)=-2.00, p=.046): there was a negative relationship between self-confidence and accuracy in participants who requested advice (slope=-1.97, SE=0.79, CI=[-3.52, -0.41]), but not for participants who declined advice (slope=2.09, SE=1.99, CI=[-1.82, 6.00]). Confidence in the GenAI was instead largely unrelated to completeness, with a main effect of advice request (B=22.67, SE=2.07, t(785)=10.98, p<.001) but no main effect of confidence in GenAI (p=.324), and no interaction (p=.713). Response completeness was thus dependent on advice-taking, with participants who requested advice being more likely to provide complete responses, perhaps due to reduced editing of the advice which was itself very detailed. And while confidence in GenAI was unrelated to response completeness, confidence in self predicted less complete responses in those who accepted advice. This suggests that participants who declined advice had some insight into their own performance in terms of response accuracy, while the availability of the GenAI output disrupts this calibration.

\subsubsection{Individual Characteristics}\label{subsubsec:results1_5}

In a series of secondary analyses, we explored the determinants of advice requests and reliance across participants, based on their attitudes towards GenAI as well as various measures of their usage habits and demographic characteristics. In general, rates of advice requests were positively related to overall beliefs about GenAI’s capabilities, as depicted in the plots included in Supplementary Information. We assessed the relative contribution of each predictor in a model predicting the percentage of advice requests. Participants’ self-reported abilities to plan events were negatively related to advice requests (B=-4.55, SE=1.26, t=-3.60, p<.001) and positively related to their beliefs in GenAI’s helpfulness in the event planning task (B=8.85, SE=2.52, t=3.52, p=.001), while there were no unique contributions of perceptions of GenAI’s capacity to plan events (p=.056), its accuracy and helpfulness more generally (p=.206), participants’ comfort relying on GenAI advice (p=.759) and working with the GenAI (p=.558), feelings of being in control during GenAI usage (p=.175), the perceived difficulty of and effort required for getting a helpful response (p=.322 and p=.148, respectively), or the importance of being able to modify the suggestions of GenAI in collaborative tasks (p=.528). On the other hand, advice reliance was largely unrelated to all these factors (all ps>.063). This suggests that requests to obtain advice are more sensitive to overall beliefs about GenAI’s capabilities, whereas reliance on the advice may depend on more momentary assessments of the advice. We note however that these ratings were provided by participants at the end of the experiment, and it remains unclear whether these accurately reflect participants’ general attitudes towards GenAI, as opposed to more fleeting impressions influenced by task completion.

As for usage habits, most participants in our sample had heard of GenAI systems prior to taking part in the study (82\%) and had used a GenAI system in the past (66\%). These participants reported using GenAI a fair amount (10\% more than once a day, 19\% once a day, 25\% once a week, 17\% once every two weeks, 7\% once a month, 22\% less than once a month). As expected, participants who used GenAI more frequently relied on the advice to a greater extent (r=0.16, p=.033). Usage frequency however was unrelated to decisions to request advice (r=0.06, p=.400) in the first place. This seemed surprising, as presumably users who report more frequent usage in daily life may also be more likely to request GenAI assistance in the current task. We speculate this may be due to the type of task employed here, as some participants reported in their debriefing that they had never thought of using AI for event planning.

Finally, rates of advice requests were stable across demographics, with a small effect of gender (males less likely than females to request advice: B=-0.09, SE=0.05, t=-2.01, p=.045; all other ts<0.32, ps>.750), age (r=-0.12, p=.084), highest level of education (F(4, 195)=2.00, p=.097), education topic (F(13, 186)=0.95, p=.504), employment sector (F(19, 180)=1.85, p=.021; no significant effects in post-hoc comparisons), and employment status (F(4, 195)=0.56, p=.689). Similarly, advice reliance was stable across demographics, with no relationship with gender (with ‘Woman’ as baseline, all ts<1.61, ps>.108), age (r=0.00, p=.952), highest level of education (F(4, 179)=1.42, p=.228), education topic (F(13, 170)=1.13, p=.338), employment sector (F(19, 164)=0.91, p=.576), and employment status (F(4, 179)=1.16, p=.328).

\subsection{Discussion}\label{subsec:discussion1}

In summary, decisions to request and rely on GenAI advice were associated with higher prospective confidence in the GenAI, and lower confidence in participants’ own abilities. These metacognitive estimates were also flexible, as they increased from pre- to post-task completion depending on whether participants relied on the advice. In particular, participants who declined advice, or those who requested but did not rely on the advice, showed a boost in self-confidence after the task, while those who requested the advice showed a boost in confidence in the GenAI. Beyond confidence, advice reliance was also associated with changes in metacognitive calibration: participants who declined advice showed a positive relationship between response completeness and confidence in oneself (suggesting they were metacognitively calibrated), while participants who relied on the advice showed no relationship between accuracy and confidence in themselves or the GenAI. Poor calibration in participants who request and rely on GenAI advice may stem from insufficient consideration of their confidence in themselves, and instead a heavier reliance on their impressions of GenAI. However, we interpret this result with caution given the overall low rates of inclusion for this omitted piece of information, and the high response completeness for participants who heavily relied on the advice.

Taken together, these results highlight an important role for confidence in working with GenAI tools, with important consequences for participants’ evaluations of their performance. In particular, the changes in confidence from prospective to retrospective measures suggest that completing a task boosts users’ own confidence in their ability, while receiving support from a GenAI might boost confidence in the capabilities of the GenAI instead. Since these measures were obtained just before and after each task, these effects may be due to the completion of the task itself, with participants’ task experience driving changes in retrospective confidence. In the current experiment, however, participants could choose whether to obtain and rely on the GenAI advice—raising the possibility that changes in confidence may be driven by the characteristics of the participants themselves, rather than exposure to advice or the completion of the tasks. We thus ran a new study where we experimentally manipulated the availability of the advice, in a direct test of the causal link between GenAI support and confidence updating.

\section{Study 2: A Causal Test}\label{sec:study2}

To further investigate the influence of confidence on advice-taking from a GenAI system, we designed a new experiment where participants were randomly assigned to complete the task in isolation or receive advice from a GenAI. We hypothesized that if advice exposure causally impacts confidence, participants would show an increase in retrospective confidence in GenAI (or in themselves) depending on whether they received (or did not receive) advice, even when they were randomly assigned to receive advice. Alternatively, if changes in confidence were dependent on the characteristics of the users themselves (rather than task completion), there would be no changes in metacognitive estimates with random assignment.

\subsection{Method}\label{subsec:method2}
The preregistered methods and analyses for this study were preregistered and can be accessed at \url{https://aspredicted.org/dkv9-dytw.pdf}, and anonymized raw data and analysis code can be viewed at \url{https://osf.io/nxpb6/?view\_only=29a813136da24784b8f45bca0a9d8bae}. The recruitment, design, and procedure for Study 2 were identical to Study 1, except as noted here. 

\subsubsection{Participants}\label{subsubsec:participants2}

This sample size was chosen and preregistered before data collection began based on a power analysis of pilot data (N=40, valid N=35), which revealed a strong interaction between prospective confidence in GenAI and advice exposure on retrospective confidence in GenAI (t(50)=4.19, p<.001). While the pilot revealed a small sample size would be sufficient to achieve 95\% power to detect this interaction (N=35), and half of this effect (N=112), we adopted a more conservative strategy and increased our target sample size to N=200. We also determined that, should a Bayesian linear mixed model for the main analysis be inconclusive based on these 200 subjects (i.e., with a BF between 0.3 and 3), we would recruit additional subjects for a total N=300 participants. Given this analysis did indeed reveal inconclusive results, we continued data collection until a total sample of 300 valid participants. Data were collected from August to October 2024.

Participants were excluded if they failed to select the correct response in an attention check at the end of the experiment (N=23), encountered problems (N=4), or failed to enter text for two or more of the four scenarios (N=7). No participants triggered the other exclusion criteria. These participants were excluded and replaced until our target sample size was reached (N=300, 152 women, 147 men, 1 non-binary, age M=46.62, SD=20.25; see Supplementary Information for more details on participant demographics). 

\subsubsection{Design}\label{subsubsec:design2}

Participants were randomly assigned to the advice condition (N=146) and the no advice condition (N=154). For participants in the advice condition, advice was provided for the middle events (2 and 3) while on events 1 and 4 there was no advice available for all participants. This allowed us to measure participants’ performance when completing the task without GenAI advice, both prior to trials where they did receive advice (event 1), as well as after exposure to advice on different events (event 4). Advice was provided in the form of bullet points, consistent with ChatGPT’s default formatting. Given the overall high accuracy in Study 1, which limited the significance of participants’ response verification metric, in this new study the completeness of GenAI advice was further reduced by removing two additional points (for details, see Supplementary Information).

\subsubsection{Procedure}\label{subsubsec:procedure2}

The wording of the questions probing participants’ confidence was slightly modified to refer specifically to the events in question, for added clarity, and the question probing confidence in GenAI usability was omitted given the high correlation with confidence in GenAI revealed in Study 1. Participants were thus asked to rate their confidence in themselves (\textit{“For the event you are about to plan, how confident are you in planning this event?”}) and in the GenAI (\textit{“For the event you are about to plan, how confident are you in GenAI systems’ information for planning this event? If you don’t have any prior experience with GenAI systems, provide your best guess.”}). The retrospective questions were modified accordingly, for both confidence in self (\textit{“For the event you planned on the previous page, how confident were you in planning this event?”}) and in the GenAI (\textit{“For the event you planned on the previous page, how confident were you in GenAI systems’ information for planning this event? If you didn’t use the GenAI system, give us your best guess imagining you had used it.”}).

\subsubsection{Overview of Analyses}\label{subsubsec:analyses2}

\paragraph{Advice-Taking} To examine the impact of advice-taking on retrospective confidence, we analysed trials involving the between-subjects manipulation of advice exposure (events 2 and 3) via: (1a) a linear regression predicting retrospective confidence in self from prospective confidence in self, advice exposure, and their interaction; and (1b) a linear regression predicting retrospective confidence in GenAI from prospective confidence in GenAI, advice exposure, and their interaction. We also tested changes in confidence based on advice reliance (rather than exposure), quantified as the similarity between participants’ responses and the advice they received, resulting in the following models: (2a) a linear regression predicting retrospective confidence in self from prospective confidence in self, advice reliance, and their interaction; and (2b) a linear regression predicting retrospective confidence in GenAI from prospective confidence in GenAI, advice reliance, and their interaction. These analyses are reported in Section \ref{subsubsec:results2_1}, \textit{From Prospective to Retrospective Confidence}.

\paragraph{Confidence Carryover} To examine the impact of advice exposure on confidence on subsequent trials, we analysed participants’ prospective confidence on trials where they were not exposed to advice (events 1 and 4). In particular, we ran: (3a) a linear regression predicting prospective confidence in self from trial number (1 and 4, i.e. pre-task and post-task, respectively), advice exposure (on trials 2 and 3), and their interaction; and (3b) a linear regression predicting prospective confidence in GenAI from trial number (1 and 4, i.e. pre-task and post-task, respectively), advice exposure (on trials 2 and 3), and their interaction. These analyses are reported in Section \ref{subsubsec:results2_2}, \textit{Carryover Changes in Confidence}.

\paragraph{Accuracy} In a set of secondary analyses, we examined whether the relationship between advice exposure and confidence was affected by response completeness, operationalized as the number of distinct steps participants mentioned in their responses, out of possible steps from a model answer (again scored via ChatGPT; see Supplementary Information). We thus re-ran models 1a-1b and 3a-3b, but now adding accuracy as a co-variate. In addition, we directly examined the role of confidence and advice exposure on accuracy in two models: (4a) a linear regression predicting response completeness from prospective confidence in self, advice exposure, and their interaction; and (4b) a linear regression predicting response completeness from prospective confidence in GenAI, advice exposure, and their interaction. In a set of additional exploratory analyses, we examined accuracy operationalised as response verification, namely the proportion of steps mentioned in participants’ responses out of the three critical steps that were removed from the advice of the GenAI. These models mirrored models 4a-4b above: (5a) a linear regression predicting the proportion of omitted information included in the response from prospective confidence in self, advice exposure, and their interaction; and (5b) a linear regression predicting the proportion of omitted information included in the response from prospective confidence in GenAI, advice exposure, and their interaction. These analyses are reported in Section \ref{subsubsec:results2_3}, \textit{Task Performance and Metacognitive Calibration}.

\paragraph{Study 1-2 Comparison} In a final analysis, we compared the primary analyses (1a and 1b) across Study 2 (where participants were randomly assigned to advice exposure) and Study 1 (where participants could request or decline advice). This resulted in the following models: (6a) a linear regression predicting retrospective confidence in self from prospective confidence in self, advice exposure, study number, and their interactions; and (6b) a linear regression predicting retrospective confidence in GenAI from prospective confidence in GenAI, advice exposure, study number, and their interactions. These analyses are reported in Section \ref{subsubsec:results2_4}, \textit{Forced vs. Requested Advice Exposure}.

\subsection{Results}\label{subsec:results2}

\subsubsection{From Prospective to Retrospective Confidence}\label{subsubsec:results2_1}

\begin{figure}[H]
\centering
 \includegraphics [width=\linewidth]{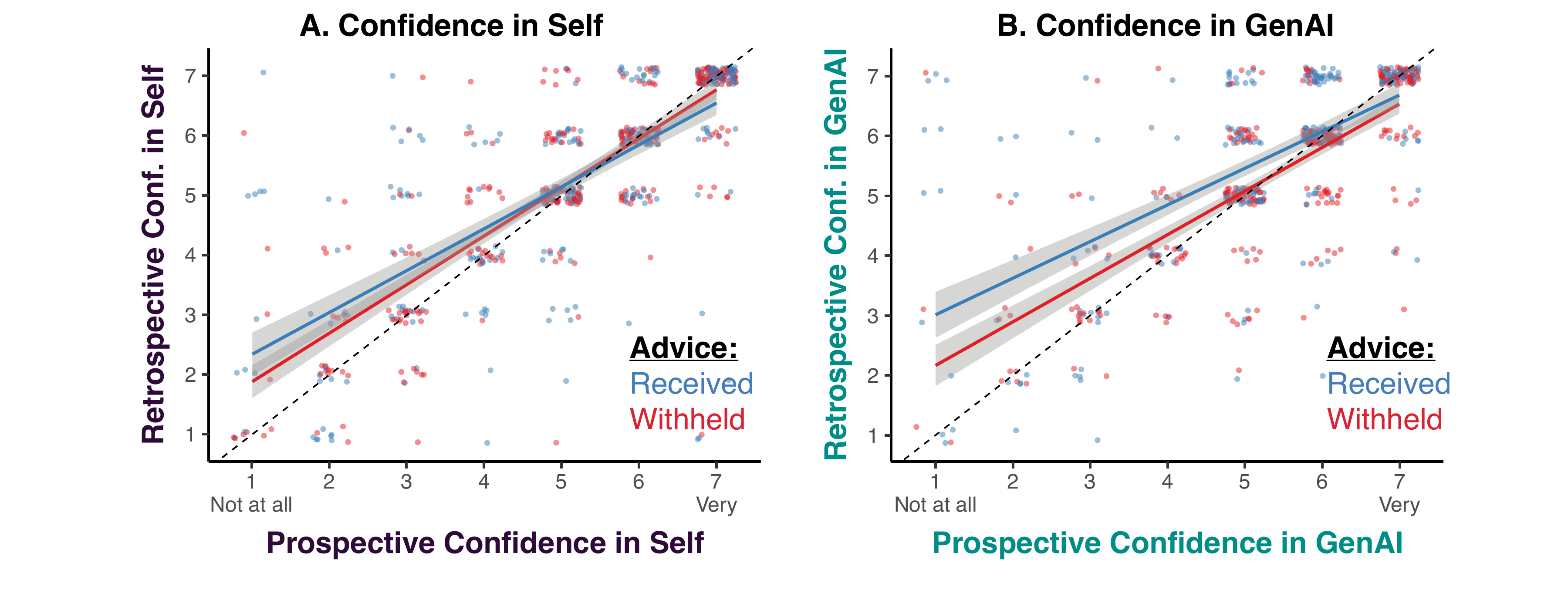}
\caption{Changes from Prospective to Retrospective Confidence in Study 2. (A) Confidence in self was overall consistent from before to after the task, but there was a greater increase from prospective to retrospective confidence in participants who were exposed to GenAI advice. (B) Confidence in GenAI was overall consistent from before to after the task, but there was a greater increase from prospective to retrospective confidence in participants who were exposed to GenAI advice. Jittered points correspond to single trials.}\label{fig_study2results1}
\end{figure}

Our primary analysis focused on how receiving advice (per the between-subjects manipulation of advice exposure) influenced changes from prospective to retrospective confidence. To find out, we ran two preregistered linear mixed models predicting retrospective confidence ratings in trials 2 and 3 from prospective confidence ratings, advice exposure, and their interaction, separately for confidence in GenAI and confidence in self. Participants who were more confident in themselves prior to the task were also more confident after the task (main effect of prospective confidence in self: B=0.79, SE=0.04, t(544)=21.12, p<.001), and there was no difference in self-confidence depending on the advice exposure condition (no main effect of advice exposure: p=.999). Importantly, there was also an interaction between these factors (B=-0.11, SE=0.05, t(531)=-2.04, p=.042; \autoref{fig_study2results1}A): participants who did not receive advice had a strong correlation between prospective and retrospective ratings (slope=0.79, SE=0.04, CI=[0.72, 0.87]), while this correlation was weaker in those who did receive advice (slope=0.68, SE=0.04, CI=[0.61, 0.76]). This shows how advice exposure causally boosts participants’ own self-confidence, which increased to a greater extent when participants were randomly assigned to receive (vs. not receive) Gen AI advice. The manipulation of advice exposure in this experiment thus revealed a different pattern than in Study 1, where boosts in self-confidence were instead greater in participants who declined (and thus were not exposed to) the advice; we examine this difference more directly in the between-study analyses reported below (Section \ref{subsubsec:results2_4}).

Participants who were more confident in the GenAI prior to the task were also more confident after the task (main effect of prospective confidence in GenAI: B=0.74, SE=0.05, t(409)=16.02, p<.001), and participants who received advice were retrospectively more confident in the information provided by the GenAI (main effect of advice taking: B=0.24, SE=0.06, t(250)=3.83, p<.001). Importantly, there was also an interaction between these factors (B=-0.14, SE=0.06, t(437)=-2.20, p=.028; \autoref{fig_study2results1}B): the correlation between prospective and retrospective ratings was stronger in participants who were randomly assigned to not receive advice (slope=0.74, SE=0.05, CI=[0.65, 0.83]) compared to those who did receive the advice (slope=0.60, SE=0.04, CI=[0.52, 0.69]). These results of confidence in GenAI thus replicated the findings of Study 1, while also showing a causal role of GenAI advice exposure in confidence boosts—such that even when randomly assigned to receive advice, participant increased their confidence in the GenAI, while these estimates remained more stable for participants who completed the task in isolation.

As for advice reliance, plans submitted after receiving advice were highly similar to the GenAI output, with 25\% of these responses showing greater than 80\% similarity with the advice (compared to only 1\% for responses where advice was not received). Analyses of reliance and confidence generally confirmed the results of advice exposure, with higher retrospective confidence in participants who were more confident prospectively (confidence in self: B=0.72, SE=0.11, t(239)=6.54, p<.001; confidence in GenAI: B=0.68, SE=0.12, t(267)=5.94, p<.001), and in participants who relied on the advice to a greater extent (confidence in self: B=0.36, SE=0.18, t(172)=2.01, p=.046; confidence in GenAI: B=0.86, SE=0.20, t(157)=4.37, p<.001). These analyses however revealed no interaction between advice reliance and prospective confidence (confidence in self: p=.719; confidence in GenAI: p=.144). Advice reliance is thus related to increases in confidence in both self and GenAI, regardless of participants’ initial confidence.

\subsubsection{Carryover Changes in Confidence}\label{subsubsec:results2_2}

In addition to changes from prospective to retrospective confidence within the same events, we also examined the impact of GenAI advice exposure on confidence in planning for different events, namely the events on trials 1 and 4, which all participants completed without advice. In particular, we conducted two preregistered linear mixed models predicting prospective confidence ratings from the trial number (1 and 4, i.e. pre-task and post-task, respectively), between-subjects condition of advice exposure (on trials 2 and 3), and their interaction, separately for confidence in self and confidence in GenAI. Self-confidence overall decreased over the course of the experiment (main effect of trial number: B=-0.06, SE=0.03, t(297)=-2.57, p=.011), but advice exposure did not impact self-confidence (no main effect of advice exposure, p=.641; and no interaction, p=.997). Confidence in GenAI did not differ based on advice exposure (no main effect of advice exposure, p=.149) or over the course of the experiment (no main effect of trial number, p=.370), with no interaction between these factors (p=.068). Thus, self-confidence declined across the experiment regardless of advice exposure, while confidence in GenAI remained stable, and advice exposure did not impact confidence on subsequent trials.

\subsubsection{Task Performance and Metacognitive Calibration}\label{subsubsec:results2_3}

As per our preregistered plan, we also re-ran the analyses reported in sections \ref{subsubsec:results2_1} and \ref{subsubsec:results2_2}, while also adding response completeness as a co-variate. On trials where advice exposure was manipulated (events 2 and 3), participants who did not receive advice provided overall less detailed plans (mean \% points included=43.50\%, SE=1.68) compared to those who did receive advice (mean score=55.50, SE=1.95). Response completeness was unrelated to retrospective confidence in self (p=.129). The remaining effects remained stable when controlling for accuracy, with a strong correlation between retrospective and prospective confidence in self (B=0.79, SE=0.04, t(544)=20.95, p<.001), especially for participants who were not exposed to the advice (slope=0.79, SE=0.04, CI=[0.71, 0.86]), compared to those who did receive advice (slope=0.68, SE=0.04, CI=[0.61, 0.76]; interaction B=-0.11, SE=0.05, t(527)=-1.98, p=.048). In contrast to confidence in self, retrospective confidence in GenAI was higher for those who provided more complete responses (B=0.27, SE=0.14, t(548)=2.00, p=.046), likely due to high reliance on the provided plan. The remaining effects remained stable when controlling for response completeness, with higher confidence in GenAI for those who received the advice (B=0.21, SE=0.07, t(265)=3.20, p=.002) and a strong correlation between retrospective and prospective confidence in GenAI (B=0.74, SE=0.05, t(410)=15.96, p<.001), especially in participants who did not receive advice (slope=0.74, SE=0.05, CI=[0.65, 0.83]), compared to those who did receive advice (slope=0.59, SE=0.04, CI=[0.51, 0.68]; interaction B=-0.15, SE=0.06, t(441)=-2.33, p=.020).

The effects of confidence also remained unaltered when controlling for response completeness in analyses of trials that all participants completed without advice (events 1 and 4). Prospective confidence in self was related to higher accuracy (main effect of response completeness: B=0.43, SE=0.17, t(562)=2.55, p=.011) and decreased over the course of the experiment (main effect of event number: B=-0.06, SE=0.03, t(301)=-2.19, p=.029), regardless of advice exposure condition (main effect p=.731; interaction p=.842). Confidence in GenAI on events completed in isolation was unrelated to response completeness (p=.296), event number (p=.305), advice exposure (p=.164), with no interaction between these factors (p=.082). While in general controlling for response completeness did not alter the effects of advice exposure and confidence, accuracy was thus related to higher prospective confidence in self, and not in GenAI.

To further probe metacognitive calibration, we directly investigated how confidence and advice exposure were related to response completeness. We ran two preregistered linear mixed models predicting response completeness in trials 2 and 3 from prospective confidence ratings, advice exposure, and their interaction, separately for confidence in GenAI and confidence in self. Participants who were more confident in themselves prior to the task provided more complete responses (main effect of prospective confidence in self: B=0.03, SE=0.01, t(585)=2.18, p=.030). Completeness also differed depending on the advice exposure condition (main effect of advice exposure: B=0.12, SE=0.02, t(295)=5.57, p<.001), with participants who received advice being more accurate than those who did not, likely due to heavy reliance on the detailed advice. Exposure to GenAI advice however did not modulate the relationship between confidence and accuracy (no interaction between confidence in self and advice exposure on accuracy, p=.154). Participants were thus overall metacognitively calibrated, with a positive relationship between accuracy and self-confidence, and GenAI advice did not impact this calibration. The analyses of confidence in GenAI revealed again a main effect of advice, with participants who received advice being more accurate (main effect of advice taking: B=0.12, SE=0.02, t(296)=5.75, p<.001); accuracy however was unrelated to confidence in GenAI (no main effect of confidence in GenAI, p=.356; no interaction, p=.241). Exposure to advice from a GenAI thus increased accuracy, regardless of participants’ confidence in themselves or in GenAI. 

In additional exploratory analyses, we examined accuracy operationalised as response verification, namely the proportion of steps participants included despite these having been removed from the advice of the GenAI (see Supplementary Information for full details on the omitted steps and the scoring procedure). Participants who were not exposed to advice were more likely to include in their plan omitted pieces of information compared to participants who were randomly assigned to receive advice (24.68 vs. 17.92\% of the omitted steps included in the final responses; X2(1)=11.80, p<.001). We next investigated the relationship between prospective confidence and response verification, in models predicting the proportion of omitted information included in the response (on events 2 and 3) from prospective confidence ratings, advice exposure, and their interaction, separately for confidence in GenAI and confidence in self. Participants who were more confident in themselves were also more likely to include omitted information (B=0.05, SE=0.02, t(594)=3.07, p=.002). Further, participants who were not exposed to advice were more likely to include the omitted information (B=-0.07, SE=0.03, t(294)=-2.62, p=.009), suggesting that participants who were exposed to the advice were not checking the output thoroughly, and thus missed key pieces of information. There was no interaction between advice exposure and self-confidence (p=.715). On the other hand, response verification was unrelated to confidence in GenAI (p=.734) and while participants who did not receive advice were again more likely to include the omitted information (B=-0.07, SE=0.03, t(296)=-2.50, p=.013), confidence in GenAI did not mediate this effect (p=.734). These effects of response verification thus confirm the analyses of response completeness, showing that participants were overall metacognitively calibrated, with a positive relationship between accuracy and self-confidence, and GenAI advice did not impact this calibration.

\subsubsection{Forced vs. Requested Advice Exposure}\label{subsubsec:results2_4}
In a final preregistered analysis, we compared the impact of advice-taking on confidence in the current experiment, where participants were randomly assigned to receive or not receive advice, to that in Study 1, where participants could choose to request or decline advice. To determine the relative impact of participants’ attitudes and advice exposure, we thus analysed changes from prospective to retrospective confidence with the type of assignment (Forced vs. Requested) as a between-subjects fixed factor. 

\begin{figure}[H]
\centering
 \includegraphics [width=\linewidth]{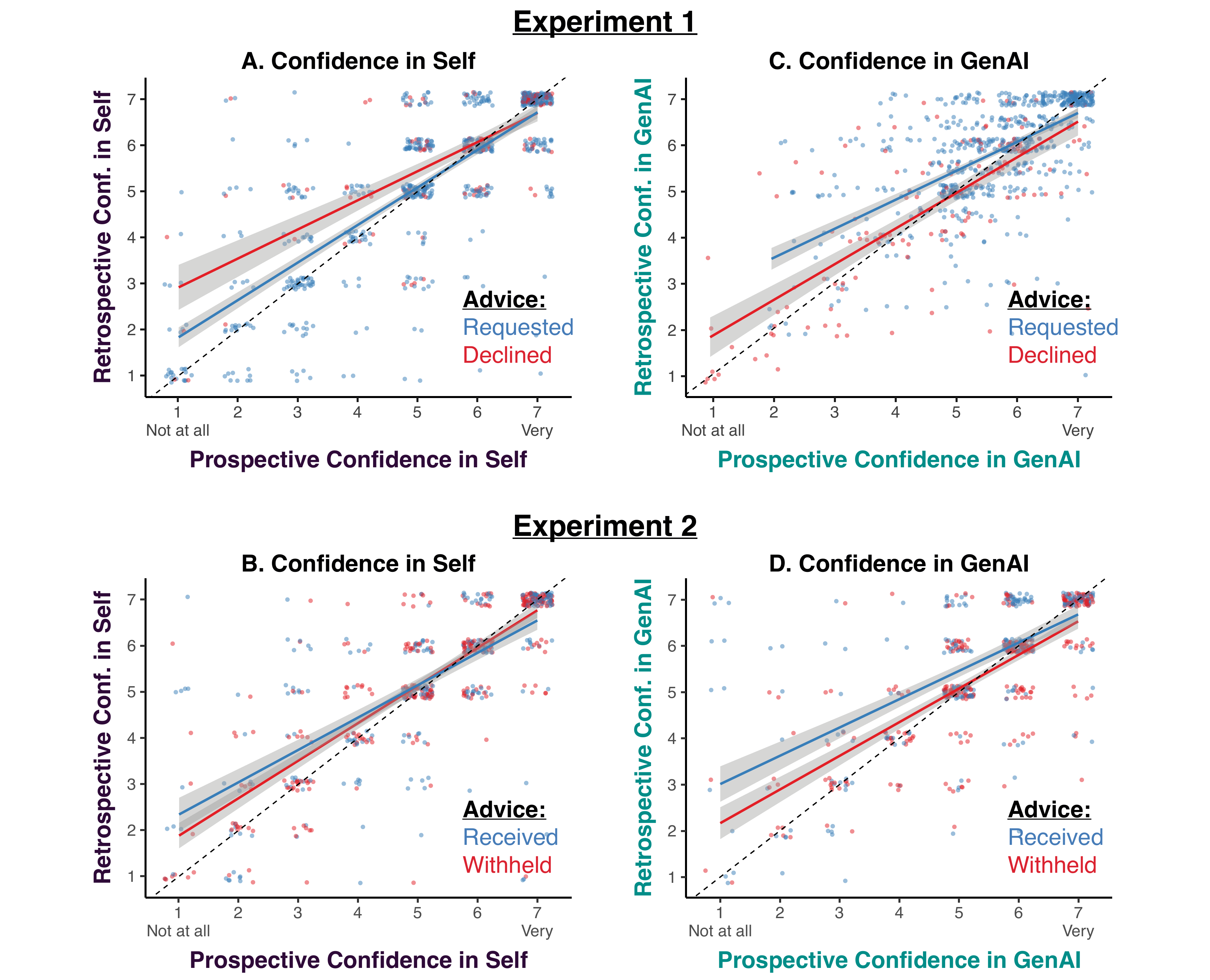}
\caption{Changes from Prospective to Retrospective Confidence in Studies 1 and 2. (A-B) For confidence in self, task completion produced an increase in retrospective confidence; this effect was particularly pronounced when participants voluntarily declined advice in Study 1, and less so when they did receive advice, or when they were randomly assigned to not receive advice in Study 2. (C-D) For confidence in GenAI, advice exposure produced an increase in retrospective confidence, both when advice exposure was voluntary (Study 1) and forced (Study 2).}\label{fig_study2results2}
\end{figure}

\paragraph{Confidence in Self} The main effects and interactions across experiments largely reflected the results of the separate analyses of Study 1 and Study 2: retrospective confidence was higher in participants who were more confident before the task (main effect of prospective confidence: B=0.42, SE=0.13, t(1271)=3.20, p=.001; \autoref{fig_study2results2}A-B), and in participants who did not request or receive advice (main effect of advice request/exposure: B=-0.38, SE=0.14, t(1197)=-2.77, p=.006). As in Study 1, there was interaction between advice exposure and prospective confidence (B=0.47, SE=0.14, t(1291)=3.31, p=.001), wherein the relationship between prospective and retrospective confidence was stronger for participants who requested or received advice (slope=0.43, SE=0.01, CI=[0.41, 0.46]) versus those who did not (slope=0.42, SE=0.02, CI=[0.38, 0.46]). 

These effects however also differed depending on whether participants could choose to request advice (Study 1; \autoref{fig_study2results2}A) or were randomly assigned to view it (Study 2; \autoref{fig_study2results2}B). First, retrospective confidence was higher for participants in the first vs. second experiment (main effect of experiment number: B=-0.17, SE=0.07, t(1297)=-2.46, p=.014). This effect also interacted with advice exposure (B=0.19, SE=0.08, t(1198)=2.30, p=.022), such that retrospective confidence was highest for those who declined advice in Study 1 (mean for scaled confidence=0.17, SE=0.07), compared to those who requested the advice in Study 1 (mean=-0.02, SE=0.05), those who received advice in Study 2 (mean=0.00, SE=0.05), and those who did not receive the advice in Study 2 (mean=0.00, SE=0.05). There was also an interaction between prospective confidence and exposure type (B=0.19, SE=0.07, t(1306)=2.71, p=.007), such that the relationship between prospective and retrospective confidence was stronger when the advice was received (Study 2; slope=0.44, SE=0.02, CI=[0.41, 0.47]), as opposed to deliberately requested (Study 1; slope=0.41, SE=0.02, CI=[0.37, 0.45]). This suggests that the highest increase in retrospective confidence, relative to prospective confidence, occurred for participants who voluntarily declined advice in Study 1.

The differential effects of prospective confidence and advice exposure in the two studies were demonstrated by a three-way interaction (B=-0.30, SE=0.08, t(1299)=-3.53, p<.001): when participants had the option to receive advice in experiment 1, those who declined advice showed an increase in self-confidence from pre-task to post-task (slope=0.36, SE=0.04, CI=[0.29, 0.43]), while these ratings remained stable for those who requested advice (slope=0.46, SE=0.01, CI=[0.43, 0.49]). Conversely, advice exposure in experiment 2 led to increases in confidence estimates from pre-task to post-task (slope=0.40, SE=0.02, CI=[0.36, 0.45]), and these ratings remained stable for those who never saw the advice (slope=0.47, SE=0.02, CI=[0.43, 0.52]).

\paragraph{Confidence in GenAI} In contrast to self-confidence, the results of confidence in Gen-AI were consistent across experiments (\autoref{fig_study2results2}C-D). As in the separate analyses of Study 1 and Study 2, there was a main effect of prospective confidence, where retrospective confidence was higher in participants who were also more confident prior to the task (B=0.78, SE=0.11, t(1308)=7.14, p<.001). There was also a main effect of advice exposure, where confidence in GenAI was higher in participants who requested or received advice (B=0.44, SE=0.16, t(1243)=2.83, p=.005), regardless of participants’ initial confidence (no interaction between advice exposure and prospective confidence, p=.249). Retrospective confidence was the same across experiments (main effect of experiment: p=.222), and so was the relationship between prospective and retrospective confidence (interaction between experiment and prospective confidence: p=.581). There was also no interaction between advice-taking and experiment number (p=.291), and no three-way interaction (p=.908).

\subsection{Discussion}\label{subsec:discussion1}

The current experiment was designed to probe the causal role of advice exposure on changes in confidence across task completion. This manipulation revealed several important insights into the metacognitive mechanisms of advice-taking from GenAI. First, participants who received advice exhibited an increase in confidence in GenAI from prospective to retrospective ratings; this provides a replication of the findings Study 1, while also highlighting an important role for advice exposure (as opposed to individual decisions to request advice) in shaping this bias, since increases in confidence now emerged when participants were randomly assigned to receive advice. On the other hand, the increase in self-confidence exhibited by participants who declined advice in Study 1 was now reversed: the correlation between prospective and retrospective confidence was now weaker in participants who did not receive advice. This suggests that boosts in self-confidence are specifically associated with decisions to decline advice, while these ratings remain more stable when the advice is not available in the first place. In addition, the current results revealed that increases in confidence in both self and GenAI were driven by advice exposure rather than advice reliance, which was related to prospective confidence but did not mediate changes from prospective to retrospective confidence. An analysis of trials that all participants completed without advice (trials 1 and 4) also revealed that changes in confidence from pre- to post-task completion did not carry over to different events, where confidence did not differ for participants who were assigned to receive or not receive advice. Finally, analyses of participants’ response completeness suggested that participants were overall calibrated, as those who were more confident in themselves provided more complete responses. Participants who received GenAI advice however also missed important pieces of information, suggesting that they were not properly checking the output. This result is consistent with the findings of Study 1, suggesting that GenAI advice increases overall accuracy yet is often not sufficiently checked. 

\section{General Discussion}\label{sec:gendiscussion}
As GenAI is increasingly integrated into users’ workflows and personal lives, key questions arise about the conditions under which users choose to rely on these tools, and how reliance in turn impacts performance and task success. GenAI tools in particular represent a novel and unique domain of external assistance, given their generalisability (in that they are suitable to a wider range of tasks) and the continuous nature of their output (as opposed to binary decisions or recommendations), raising important questions about how users integrate this information. In particular, we considered that the usage of external information on a given task depends on users’ assessments of their own abilities (confidence in self) as well as their beliefs about the accuracy of the system (confidence in GenAI). To investigate cognitive mechanisms of reliance on GenAI advice, we designed a novel task where participants formulated plans for different types of events and were either given the opportunity or randomly assigned to receive advice from a GenAI system (in studies 1 and 2, respectively). This allowed us to investigate decisions to request advice, but also the extent of reliance—closely resembling the unique yet increasingly common types of decisions users face when using GenAI. 

The results of the current investigations highlight a key role for confidence in both decisions to seek out and rely on the advice of GenAI. First, decisions to request advice depended on participants’ confidence in the system, with more requests from those who were more confident in GenAI’s capabilities \citep[in line with work on non-GenAI systems;][]{klingbeil_trust_2024,yin_understanding_2019,kahr_understanding_2024}. However, decisions to request advice also depended on participants’ own confidence, with more requests from those who were less confident in themselves. These effects align with studies on non-generative tasks and AI \citep{jessupCloserLookHow2024,dreiseitl_physicians_2005} or other automated systems \citep{lee_trust_1994}. They also replicate past work in human-human interactions, where reliance on advisors increases when advice-takers are more confident in the advisors’ abilities \citep[e.g.,][]{carlebach_flexible_2023} and when they are less confident in their own abilities \citep[e.g.,][]{pescetelli_confidence_2021}. But beyond guiding advice requests, confidence in self and GenAI also played a role in advice reliance, measured as the textual similarity between GenAI output and participants’ final responses: even among participants who requested advice, greater reliance was related to lower confidence in themselves and greater confidence in the system. Confidence thus plays a role not just in deciding whether to take GenAI advice, but also in how the advice is incorporated into a final decision.

In contrast to our findings, \citet{chong_human_2022} found that primarily self-confidence, rather than confidence in AI, influences AI reliance (in this case, accepting chess move suggestions from a non-generative AI system). Yet these latter findings stand in contrast to earlier work on automation which, like the current study, found that both confidence in self and confidence in automation is important \citep{madhavan_similarities_2007,lee_trust_1994}. The complexity of the task and/or system output may be key factors explaining this difference. Our event planning task involves multiple factors that participants need to consider and integrate into their final response (e.g., attendees, location, timing etc.); the GenAI output participants received as advice reflects equal complexity. Similarly, \citet{lee_trust_1994} studied a simulated pasteurization plant, where participants must consider multiple dynamic parameters and controls that may be under manual or automatic control, as well as multiple objectives (e.g., performance, safety etc.). By contrast, \citet{chong_human_2022} use chess, which, although complex, has a single objective and progresses in turns, and participants experienced an AI system that suggested discrete moves without any further reasoning, leaving participants with little information from the AI to inform confidence estimates. Despite GenAI being deployed for complex, generative, and often subjective tasks, such as writing and coding, many studies on GenAI reliance continue to use simplified and discrete decision-making tasks for experimental control \citep[e.g.,][]{kim_im_2024,ma_are_2024,bo_rely_2024}. As an example of why this matters, consider that the impact of verbal qualifiers in GenAI responses on AI reliance varies substantially by subject domain \citep{zhou_rel-i_2024}. We therefore suggest that task and/or AI output complexity are under-explored factors to consider in future research on metacognition and AI reliance more broadly. 

Beyond the influence of prospective confidence on advice reliance, our results also indicated that exposure to GenAI in turn causally increased retrospective confidence in participants’ responses. While prospective and retrospective confidence were overall strongly correlated, in both studies there was a marked increase in retrospective confidence after task completion, both with respect to confidence in self and in GenAI. The increase in self-confidence when participants were randomly assigned to receive GenAI advice in Study 2 is consistent with past studies of web search behaviour, where participants report an increase in self-assessed knowledge following searches, misattributing retrieved knowledge as their own \citep{fisher_searching_2015,dunn_distributed_2021}. A similar misattribution effect was observed in a decision support study with a non-generative AI system \citep{chong_human_2022}, and in a recent study of GenAI support for logical problem-solving, where reliance on GenAI advice led participants to becoming over-confident in their own performance \citep{fernandes_ai_2024}. This misattribution effect implies a risk that GenAI tools may warp our confidence calibration, leading to inappropriate AI reliance.

Additionally, we observed increases in retrospective self-confidence among participants who were not exposed to GenAI advice, and these increases were stronger in participants who voluntarily declined advice in Study 1, compared to those who were randomly assigned to not receive advice in Study 2. This suggests that self-confidence not only guides decisions to accept or decline GenAI advice, but that this decision in turn reinforces participants’ own confidence, potentially consistent with a type of choice-selective bias \citep{lind_choice-supportive_2017}. The effect of advice exposure on self-confidence in Study 1 was also clear in analyses of advice reliance, as participants who initially requested yet later did not rely on the advice showed greater boosts in self-confidence. This suggests that the reinforcing effect of advice rejection on self-confidence persists beyond initial decisions and throughout the task, as participants consider whether to integrate the GenAI output they receive. This increase in self-confidence after declining advice may be driven by the evidence that participants could indeed complete the task without advice. Alternatively, it may reflect a motivation to justify the declining of advice, as if participants were motivated to rationalize their decision by inflating their confidence afterward. The latter explanation is also consistent with the lack of evidence for carry-over across trials, reflecting a framing of previous decisions rather than an updating of global beliefs. Future work may further probe these effects and their sources by eliciting participants' own explanations and assessments of specific aspects of their performance.

The bidirectional relationship between confidence and GenAI advice exposure also extended to confidence in GenAI, as confidence in the system increased when advice exposure was voluntary (as per participants’ requests in Study 1) and when forced (as per random assignment in Study 2). That advice exposure increased retrospective confidence in GenAI shows that participants did not fully misattribute learnt knowledge to themselves, and beyond integrating GenAI output into their own responses, they also appropriately attributed credit to the system. 

This raises important questions for future research about what aspects of the GenAI output contribute to increases in self-confidence versus confidence in the AI system. Prior research finds that people can update their self-confidence based on misleading cues in their interaction with technology, such as relying on internet search retrieval speed to update their confidence in their own subsequent retrieval of that information \citep{stone_search_2021}. Even the mere belief that one is being supported by technology such as AI can lead to ‘placebo effects’, where people increase risk-taking behaviour \citep{villa_placebo_2023} or otherwise modify their performance expectations \citep{kosch_placebo_2023,kloft_ai_2024}. Identifying and understanding the nature of such effects in GenAI is important for supporting appropriate reliance on this technology \citep{tankelevitch_metacognitive_2024,klingbeil_trust_2024}. For example, the presence of greetings in GenAI responses that signify warmth or competence increases AI reliance \citep{zhou_rel-i_2024}. The novel generative task we developed affords explorations of how participants attribute credit for different aspects of the responses which are difficult to capture in single binary decisions; these include not only the complexity of responses (e.g., containing several steps), but also a diverse set of skills (e.g., including the appropriate steps in the formulation of the plan, providing details for each step, and writing clearly and concisely).

On the other hand, increases in retrospective confidence were specific to assessment of performance in the current trial (i.e., planning event scenario), and did not carry over to confidence on different trials. This was clear in the analyses of the first and final trials of Study 2, where participants showed stable estimates regardless of whether they received GenAI advice in the intervening trials. This result suggests that, despite increases in confidence for current tasks, when it comes to different contexts, participants may assess their confidence based on prior beliefs about their own ability rather than recent task performance. This limited impact of advice exposure on global confidence may be due to the variability of tasks within the current experiment: while all trials involved event planning, the events themselves concerned disparate domains, from a camping trip to work recruitment. While the current study included multiple domains simply to ensure the generality of the results, future work may more systematically vary the similarities between different tasks to further explore the conditions under which confidence boosts may remain constrained to single trials as opposed to generalizing across tasks \citep[see also][]{fernandes_ai_2024}. In fact, the novel generative task we introduce here uniquely affords this possibility, given the range of tasks involving GenAI assistance as compared to other forms of AI.

The limited carryover of confidence on future trials result may also be due to the relatively limited number of trials in the current study, as participants received advice on two trials only. Exposure to advice on a longer timescale may reveal stronger updating effects, especially when the advice is less accurate than participants’ prior expectations \citep[see also][]{colombatto_illusions_2023}. An interesting question in this respect is how participants’ confidence in and reliance on GenAI may evolve as users gain more experience with the system, especially when they are exposed to errors \citep{chong_human_2022}. While rates of response verification in the current experiment were low, suggesting participants did not realise information was omitted from the output, future work may more systematically vary the quality of GenAI output to explore its effect on confidence in and reliance on GenAI.

The influence of participants’ confidence in their own abilities and in the GenAI capabilities on advice requests was clear not just in participants’ ratings on each trial, but also in the overall assessments they reported at the end of the experiments. In particular, advice requests throughout the experiment were negatively predicted by participants’ self-reported abilities to plan events and positively related to their beliefs in GenAI’s helpfulness in the event planning task. We note however that these ratings were elicited for all participants only at the end of the experiment, and thus it remains unclear how overall beliefs may influence behaviour on each trial, as opposed to experience influencing post-experiment assessments. And in contrast to advice requests, these overall assessments were unrelated to advice reliance. This suggests that requests to obtain advice are more sensitive to beliefs about GenAI’s capabilities on a given task \citep[which can be inaccurate or biased, e.g., see][]{wang_investigating_2024}, whereas reliance on the advice may depend on more momentary assessments of the advice. 

Advice requests were also related to other self-reported factors, such as general perceptions of GenAI’s capabilities beyond the specific task in question \citep[see also][]{said_artificial_2023}, and participants’ comfort relying on GenAI advice (Supplementary Information). These variables, however, did not significantly predict advice requests when tested in a model including confidence in self and GenAI, suggesting that much of their predictive power is ultimately captured by participants’ confidence in their own and GenAI’s abilities. The relatively minor contribution of confidence in usability may in part be due to the design of the current tasks, where participants were provided with fixed prompts rather than being able to design or modify the prompts themselves. An interesting future direction could be to examine how users’ prompting ability, and their confidence in their prompting ability, contribute to the accuracy and confidence of their responses. 

On the other hand, advice reliance (but not requests) was highly correlated with self-reported GenAI usage frequency outside of the study context. Anecdotally, many participants reported in post-experiment debriefing that they had not previously considered using GenAI for event planning. This suggests that domain-specific usage habits may influence participants’ decisions to seek out (vs. decline) advice, such that users who use GenAI more frequently in daily life are not more likely to request GenAI assistance in new tasks \citep[see also][]{jessupCloserLookHow2024,passi_appropriate_2024}. Yet these participants may be more willing to rely on the advice once obtained. We also note that, while our sample was representative of the U.S. population in terms of demographics (age and gender), these participants were recruited via an online crowdsourcing platform. As a result, our sample may be skewed in terms of familiarity with technology and computers, and indeed most participants in our sample had heard of GenAI systems prior to taking part in the study (82\%) and had used a GenAI system in the past (65\%). 

Another important aspect of the current results concerns the accuracy of participants’ responses, which was measured both in terms of response completeness (i.e., the number of steps they mentioned relative to an ‘ideal’ response), and response verification (i.e., the number of key steps they mentioned out of those that, unbeknownst to them, were omitted from the GenAI output). Participants who requested or received advice overall provided more complete responses, often due to copying the GenAI output verbatim; indeed, respectively 53\% and 25\% of responses in Study 1 and 2 showed a cosine similarity above 0.80. That twice as many responses were copied verbatim in the voluntary (Study 1) than the forced context (Study 2) suggests that people who voluntarily seek out GenAI advice are more likely to cognitively offload the entire task to GenAI, at least in our study context. Most importantly, participants exposed to GenAI advice in both studies did not verify the GenAI output sufficiently, submitting responses that were missing important details when these were omitted from the output—even when monetarily incentivised to provide high-quality responses (i.e., we offered a bonus reward for the best overall response). Participants who did not request or receive advice instead provided less detailed responses, but were more likely to include this key information. This result is consistent with evidence that the availability of AI assistance reduces critical thinking for a given task \citep{fernandes_ai_2024,qian_take_2024,siLargeLanguageModels2024} or the effort people invest in it \citep{lee_impact_2025}. It also exposes a trade-off between the overall level of detail in a response (furnished by the verbosity of many GenAI systems) and response verification in cases where the output may contain inaccuracies. 

The effects of GenAI support on accuracy were also largely independent of participants’ confidence: while participants were overall metacognitively calibrated, with higher self-confidence on trials where they also provided more complete responses, assessments of confidence in themselves (or the system) did not predict higher rates of response verification. This is inconsistent with \citet{lee_impact_2025}, who found that, in people’s daily work tasks, higher self-confidence in a given task is associated with increased self-reported effort in critical thinking around GenAI. It is possible that the artificial task context here may have not warranted sufficient response verification, leading to the overall low verification rates. Another explanation of this discrepancy may be the participant samples: our sample varied substantially in their frequency of GenAI use, whereas \citet{lee_impact_2025} recruited participants who had used GenAI at least weekly and therefore may have developed more experience with response verification and critical thinking around GenAI. 

The relationship between confidence and accuracy was reduced in participants who requested advice in Study 1, suggesting that obtaining advice from GenAI disrupts metacognitive calibration \citep[see also][]{fernandes_ai_2024,bo_rely_2024}. However, advice exposure did not impact metacognitive calibration in Study 2, suggesting that these effects are driven by participants’ own decisions to rely on the advice rather than mere exposure to GenAI. We also note that calibration in the current studies was operationalized as a correlation between response completeness and self-reported confidence across trials and participants, given that each participant completed only four events. However, a greater number of trials per participant would allow for a more precise estimation of calibration within-subjects, based on trial-by-trial variation in accuracy and confidence. Future experiments could thus leverage the novel task we developed to allow for a more precise estimation of calibration, for example in a longer experiment or shorter generative tasks.

Our findings underscore the importance of developing interventions to help people become better calibrated in their confidence in themselves and GenAI, and facilitate appropriate reliance on the technology, including verifying responses when relevant \citep{tankelevitch_metacognitive_2024,passi_appropriate_2024}. Interventions to improve calibration of confidence in AI and thereby reduce inappropriate reliance have included providing background information on AI systems \citep{goyalWhatElseNeed2023}, explanations \citep{chen_machine_2023,vasconcelos_explanations_2023,schemmer_appropriate_2023}, natural language uncertainty expressions \citep{kim_im_2024}, uncertainty highlighting, disclaimers, and other approaches \citep{bo_rely_2024,passi_appropriate_2024}. Overall, such studies suggest that achieving appropriate reliance is challenging, with some interventions backfiring or shifting users from over- to under- reliance \citep[e.g.,][]{bo_rely_2024}. The potential of targeting self-confidence calibration to facilitate appropriate reliance has been relatively underexplored. \citet{ma_are_2024} found that interventions to calibrate self-confidence (e.g., considering counterfactuals, or thinking in bets) can increase appropriate reliance in discrete decision-making tasks but are insufficient on their own. More research is needed to explore this avenue, including in combination with other approaches, and particularly for GenAI.

Overall, the current studies present a novel task to investigate participants’ performance in generative tasks, allowing for the study of the determinants and consequences of reliance on GenAI in complex and subjective tasks. These investigations highlight a key role for confidence in shaping advice requests as well as advice reliance, consistent with advice taking in other domains; critically, advice exposure also impacted retrospective confidence in turn—demonstrating that advice causally boosts confidence in both self and GenAI, all the while people fail to sufficiently engage with and verify the advice for their tasks. These methodological and theoretical advances thus shed light on the cognitive mechanisms underpinning users’ reliance on the increasingly common yet unique forms of GenAI assistance.

\pagebreak

\section*{Declarations}

\textbf{Funding:} This work was supported by a Microsoft Research grant under the AI, Cognition, and the Economy (AICE) program, and has benefited from the Microsoft Accelerating Foundation Models Research (AFMR) grant program.

\textbf{Acknowledgments:} The authors wish to thank Roy Zimmermann and members of the Tools for Thought group at Microsoft Research for helpful discussions.

\textbf{Declaration of Competing Interests:} \textbf{S. Rintel} and \textbf{L. Tankelevitch} are employed by Microsoft which produces Generative AI tools.

\textbf{CRediT Authorship Contribution Statement:} \textbf{C. Colombatto}: Conceptualization, Methodology, Funding acquisition, Software, Investigation, Data Curation, Formal analysis, Visualization, Writing - Original Draft, Writing - Review \& Editing; \textbf{S. Rintel}: Conceptualization, Methodology, Funding acquisition, Writing - Review \& Editing; \textbf{L. Tankelevitch}: Conceptualization, Methodology, Funding acquisition, Writing - Original Draft, Writing - Review \& Editing.

\textbf{Ethics Statement:} All experimental methods and procedures were approved by a University of Waterloo Research Ethics Board (Protocol \#46224), and all participants provided informed consent.

\textbf{Data Availability Statement:} Anonymised raw data are openly available on the Open Science Framework (OSF) website at this link: \url{https://osf.io/nxpb6/?view_only=29a813136da24784b8f45bca0a9d8bae}. 

\textbf{Code Availability Statement:} Analysis code is openly available on the OSF website at this link: \url{https://osf.io/nxpb6/?view_only=29a813136da24784b8f45bca0a9d8bae}.

\pagebreak










\appendix
\section{Event Planning Prompts}
\label{app:eventprompts}

\subsection{Event 1: Team-Building Retreat}

\begin{itemize}
    \item \textbf{Instructions to Participant:} “Imagine you need to organise a one-day retreat with your co-workers at a school that you work in to foster team-building. Which steps would you take to plan this event?”
    
    \item \textbf{Prompt to LLM:} “Which steps would you advise taking to organise a one-day retreat with your co-workers at a school that you work in to foster team-building? Answer in bullepoints.” – “Now give the same answer but in prose, without bulletpoints.”
    
    \item \textbf{LLM Advice:} https://chat.openai.com/share/4a1251d2-fd8a-41da-ad7e-b0c5302cf8fd
    
    \item \textbf{Editing in Study 1:} Removed information about Catering and Refreshments (point \#8)
    
    \item \textbf{Editing in Study 2:} Removed information about Creating a Budget (point \#3) and Logistics and Transportation (point \#7)
\end{itemize}

\subsection{Event 2: Office Recruitment Event}
\begin{itemize}
    \item \textbf{Instructions to Participant:} “Imagine you need to organise an open house at your office workplace to attract and engage potential job candidates. Which steps would you take to plan this event?”
    \item \textbf{Prompt to LLM:} “Which steps would you advise taking to organise an open house at your office workplace to attract and engage potential job candidates? Answer in bullepoints.” – “Now give the same answer but in prose, without bulletpoints.”
    \item LLM Advice: \url{https://chat.openai.com/share/f57a4bae-2c34-4bb8-89b6-22e26fa04684}
    \item \textbf{Editing in Study 1:} Removed information about Prepare Presentations (point \#7)
    \item \textbf{Editing in Study 2:} Removed information about Preparing the Workspace (point \#4) and Promoting Employee Involvement (point \#6)
\end{itemize}

\subsection{Event 3: Weekend Camping Trip}
\begin{itemize}
    \item \textbf{Instructions to Participant:} “Imagine you need to organise an outdoor camping trip for the weekend with a couple of friends. Which steps would you take to plan this event?”
    \item \textbf{Prompt to LLM:} “Which steps would you advise taking to organise an outdoor camping trip for the weekend with a couple of friends? Answer in bullepoints.” – “Now give the same answer but in prose, without bulletpoints.”
    \item \textbf{LLM Advice:} \url{https://chat.openai.com/share/43bcc51a-07b4-48bd-bc3d-cb93b6235b03}
    \item \textbf{Editing in Study 1:} Removed information about Organizing Transportation (point \#8)
    \item \textbf{Editing in Study 2:} Removed information about Reserving Campsites (point \#7) and Preparing Meals (point \#9)

\end{itemize}

\subsection{Event 4: Dinner Party}
\begin{itemize}
    
    \item  \textbf{Instructions to Participant:} “Imagine you need to organise a dinner at your house with some close friends. Which steps would you take to plan this event?”
    \item \textbf{Prompt to LLM:} “Which steps would you advise taking to organise a dinner at your house with some close friends? Answer in bullepoints.” – “Now give the same answer but in prose, without bulletpoints.”
    \item  \textbf{LLM Advice:} \url{https://chatgpt.com/share/01c64901-296c-4259-b8a9-818b3f24f192}
    \item  \textbf{Editing in Study 1:} Removed information about Sending Invitations (point \#3)
    \item \textbf{Editing in Study 2:} Removed information about Preparing Your Home (point \#5) and Music and Ambiance (point \#7)

\end{itemize}

\section{Scoring Prompts}
\label{app:scoringprompts}
\subsection{Scoring Prompts for Missing Items}
\begin{itemize}
\item  \textbf{Step 1:} “Here is a plan for organizing a [camping] event: [participant’s plan]. Does this plan explicitly mention [a plan for how everyone will get to the camping site (e.g., carpooling or rental), and confirming everyone has a way of getting there?] Respond with 0 if no, 1 if yes.” 

\item  \textbf{Step 2:} “Copy the text from the plan that corresponds to that information.” 

\item  \textbf{Step 3:} “Please indicate your final response here (1 if mentioned explicitly, 0 if not).” 

\end{itemize}

\subsection{Scoring Prompts for Completeness Score}
\begin{itemize}
\item  \textbf{Step 1:} “Below are two plans for organizing a [camping] event. For each of the [20] steps mentioned Plan 1, find whether a similar step is included in Plan 2. If it is included, tell me how similar the two steps are, on a scale from 1 (not similar at all) to 5 (very similar). If it is not included, assign a score of 0. Plan 1: [participant’s plan]. Plan 2: [model plan].” 

\item  \textbf{Step 2:} “Based on your previous answer, count how many of the [20] steps from Plan 2 are also mentioned in Plan 1. A step counts as mentioned if the similarity score you assigned on the previous step is above 3 (out of 5). Respond with a number representing the number of steps explicitly mentioned in both plans. Also respond with a list of similarity scores for each step. For example, the output should look like: Number\_of\_steps: 8; Similarity\_scores: ‘0,1,0,0,1,2,3,3,2,4,4,5,5,4,5,0,2,2,4,5’.” 

\end{itemize}

\section{Sample Demographics and Self-Reported Attitudes}
\label{app:eventprompts}

\begin{table}[H]
\footnotesize
\centering
\begin{tabular}{p{0.3\textwidth} p{0.3\textwidth} p{0.3\textwidth}}
  \textbf{Variable} & \textbf{Study 1} & \textbf{Study 2} \\ \hline 
  N & 200 & 300 \\ \hline
 Age & Mean=46.55, SD=18.46  & Mean=46.62, SD=20.25 \\  \hline
Gender & Woman 100 & Woman 152 \\
 & Man 94 & Man 147 \\
 & Non-binary/gender-diverse 5 & Non-binary/gender-diverse 1 \\
 & Prefer not to answer 1 & Prefer not to answer 0  \\ \hline

Education (highest level) & High school diploma 28 & High school diploma 65  \\
 & Secondary education 0 & Secondary education 1 \\
 & Technical/community college 31 & Technical/community college 59  \\
 & Undergraduate degree 83 & Undergraduate degree 106  \\
 & Graduate degree 49 & Graduate degree 53  \\
 & Doctorate degree 9 & Doctorate degree 16   \\ \hline

Employment status & Full-Time 104 & Full-Time 130 \\
 & Not in paid work 33 & Not in paid work 50 \\
 & Part-Time 41 & Part-Time 76  \\
 & Other 11 & Other 13 \\
 & Unemployed 11 & Unemployed 31 \\ \hline

Heard of GenAI  & Yes 165 & Yes 217  \\
 & No 35 & No 83 \\ \hline

Used GenAI  & Yes 131 & Yes 166  \\
 & No 69 & No 134 \\ \hline

Usage frequency & More than once a day 13 & More than once a day 26  \\
 & About once a day 25 & About once a day 39  \\
 & About once a week 33 & About once a week 32  \\
 & About once every two weeks 22 & About once every two weeks 21  \\
 & About once a month 9 & About once a month 22  \\
 & Less than once a month 29 & Less than once a month 26  \\ \hline

\end{tabular}
\end{table}
\renewcommand{\arraystretch}{1}

\subsection{Beliefs about GenAI in Study 1}
\begin{figure}[H]
\centering
\includegraphics [width=\linewidth]{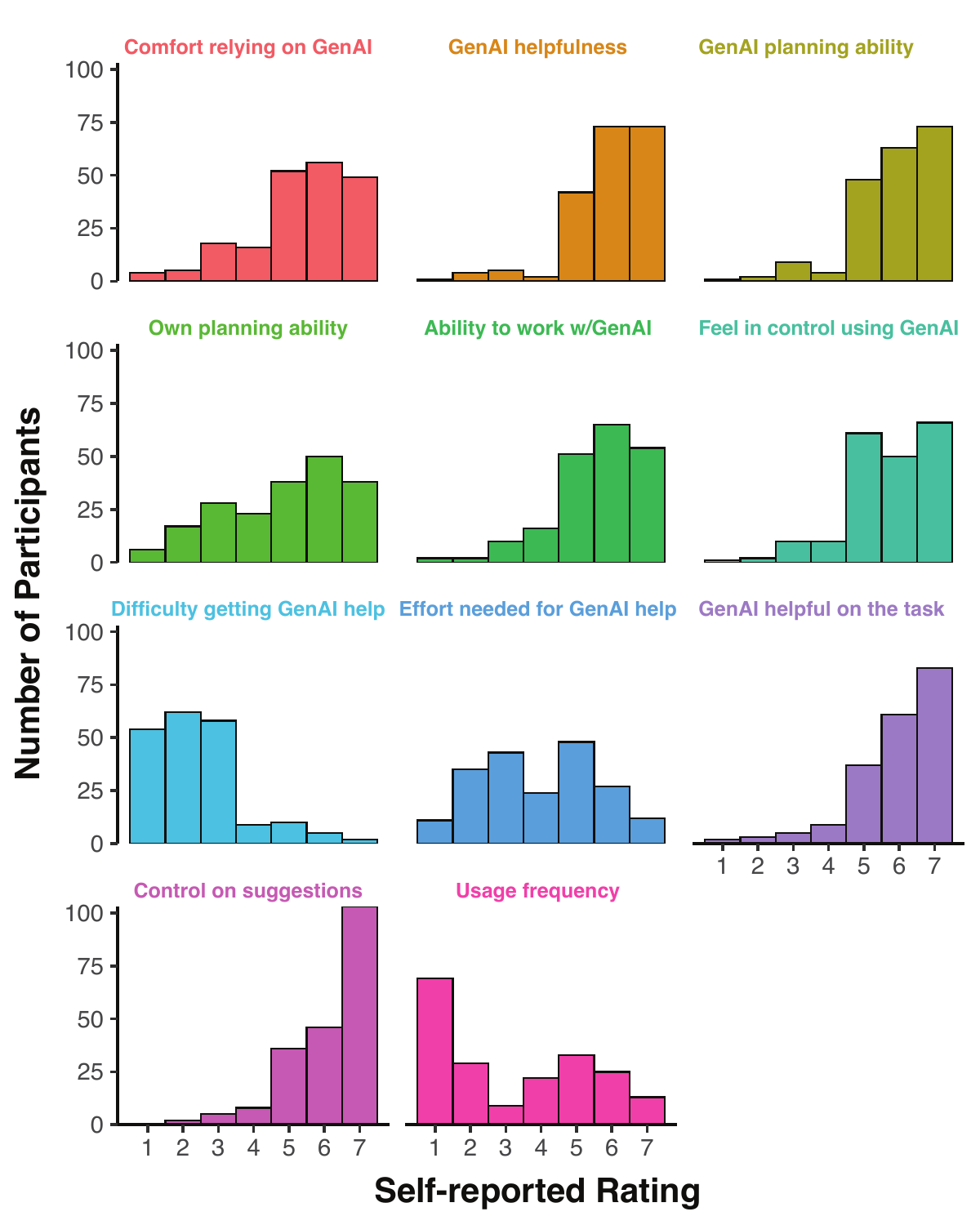}
\end{figure}

\subsection{Beliefs about GenAI in Study 2}
\begin{figure}[H]
\centering
\includegraphics [width=\linewidth]{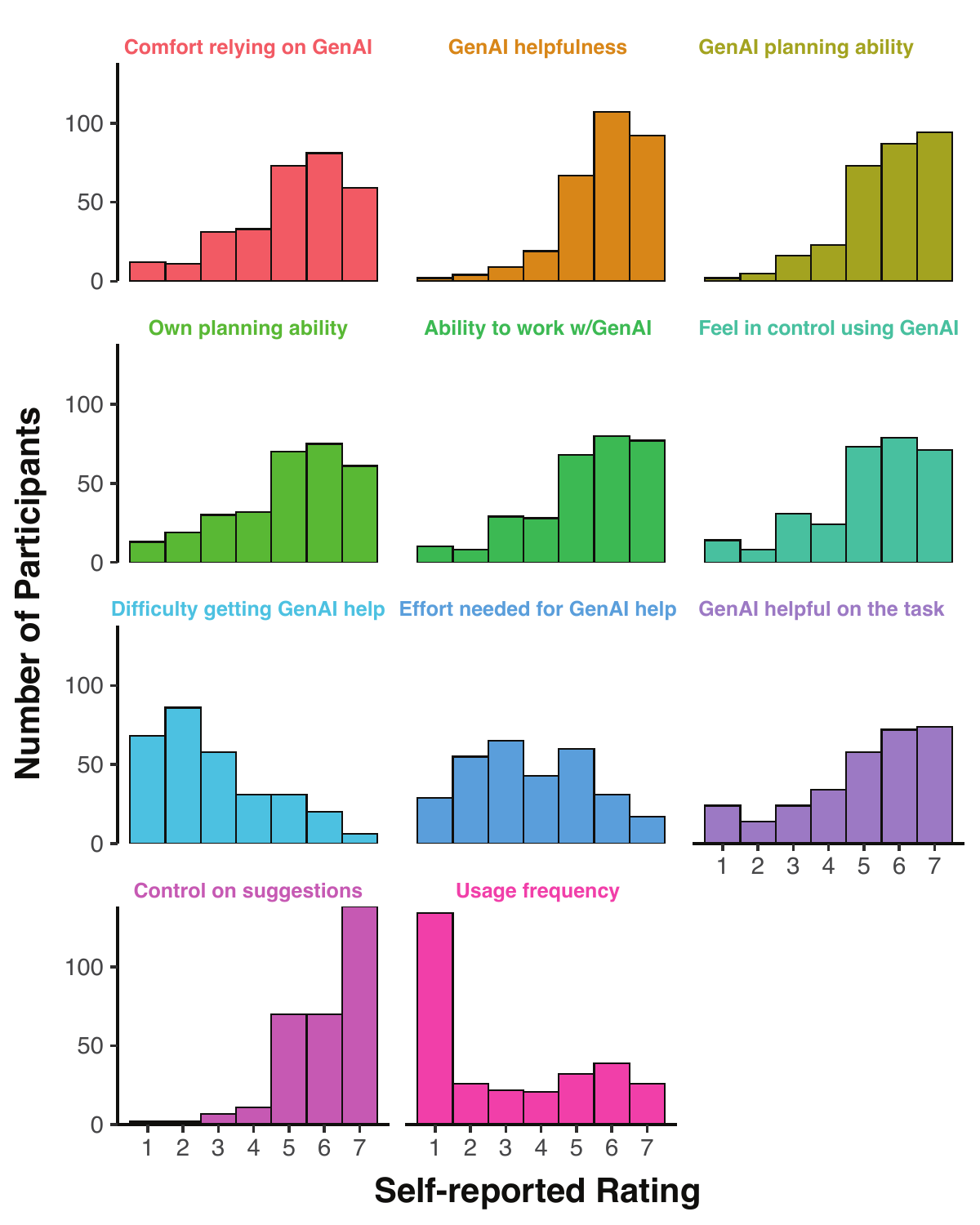}
\end{figure}

\subsection{Beliefs about GenAI and Advice Requests in Study 1}
\begin{figure}[H]
\centering
\includegraphics [width=\linewidth]{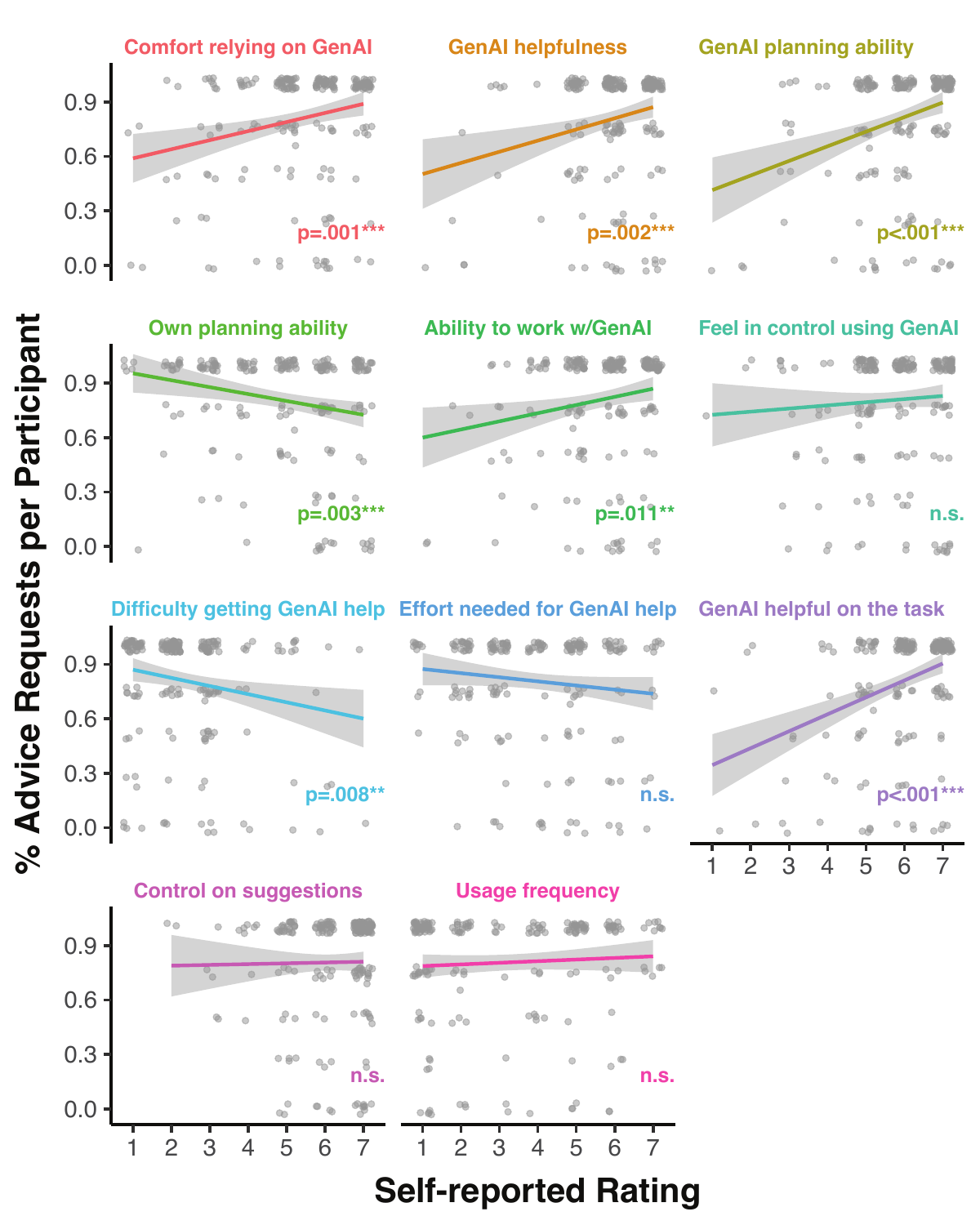}
\end{figure}

\subsection{Beliefs about GenAI and Advice Reliance in Study 1}
\begin{figure}[H]
\centering
\includegraphics [width=\linewidth]{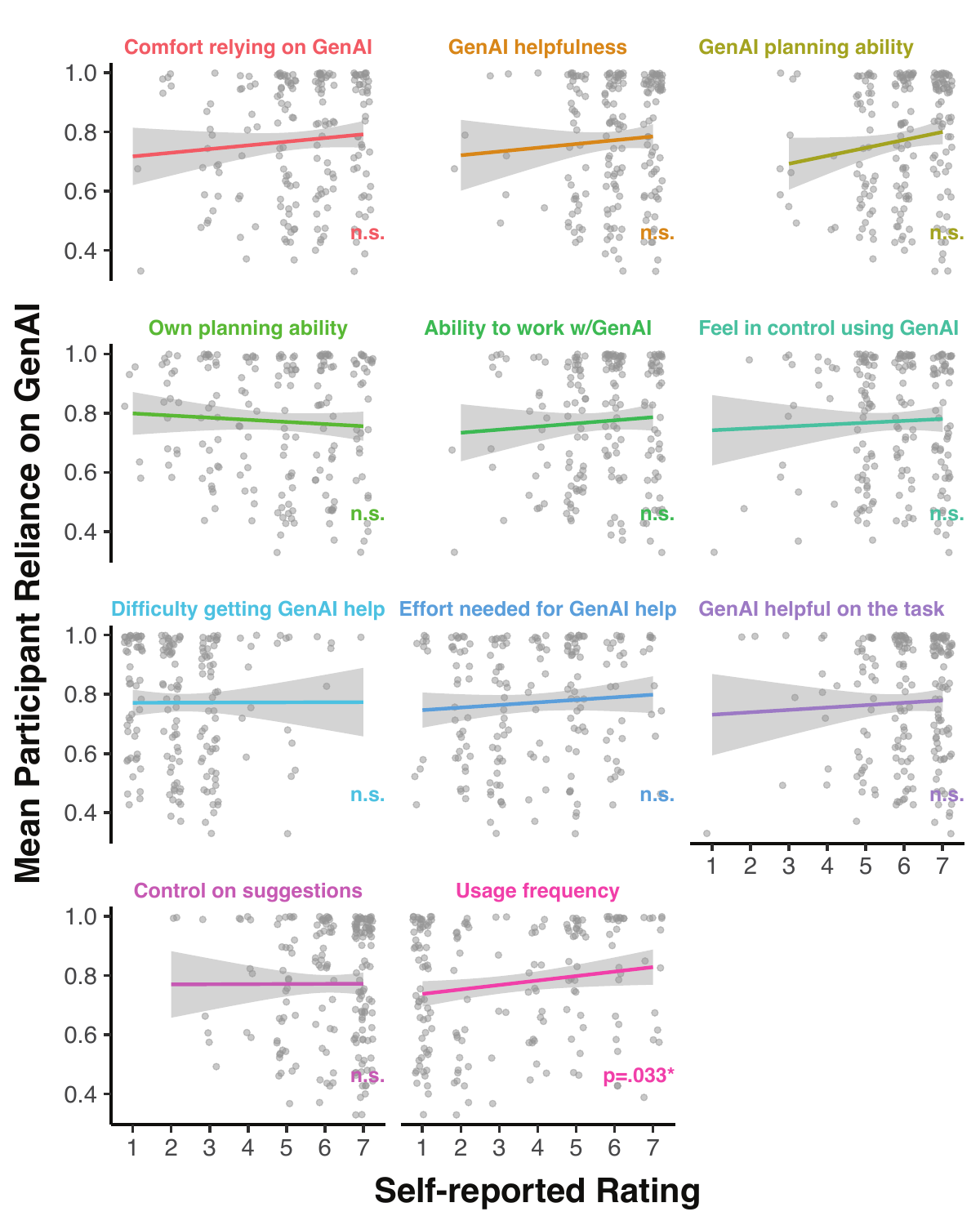}
\end{figure}

\bibliographystyle{elsarticle-harv} 
\bibliography{references}

\end{document}